\documentclass[english]{emulateapj}
\usepackage[T1]{fontenc}
\usepackage[latin9]{inputenc}
\usepackage{graphicx}
\usepackage{amssymb}

\makeatletter

\providecommand{\tabularnewline}{\\}

\shorttitle{X-ray Bursts with Short Recurrence Times}
\shortauthors{Keek, Galloway, In 't Zand, Heger}

\makeatother

\usepackage{babel}

\begin{document}

\title{Multi-Instrument X-ray Observations of Thermonuclear Bursts with
Short Recurrence Times}

\author{L.~Keek\altaffilmark{1}, D.\,K.~Galloway\altaffilmark{2}, J.\,J.\,M.~in~'t~Zand\altaffilmark{3},
A.~Heger\altaffilmark{1}}

\email{laurens@physics.umn.edu}

\altaffiltext{1}{School of Physics and Astronomy, University of Minnesota, 116
Church ST SE, Minneapolis, MN 55455, USA}

\altaffiltext{2}{Center for Stellar and Planetary Astrophysics, Monash University,
VIC 3800, Australia}

\altaffiltext{3}{SRON Netherlands Institute for Space Research, Sorbonnelaan 2,
NL - 3584 CA Utrecht, The Netherlands}
\begin{abstract}
Type I X-ray bursts from low-mass X-ray binaries result from a thermonuclear
runaway in the material accreted onto the neutron star. Although typical
recurrence times are a few hours, consistent with theoretical ignition
model predictions, there are also observations of bursts occurring
as promptly as ten minutes or less after the previous event. We present
a comprehensive assessment of this phenomenon using a catalog of 3387
bursts observed with the BeppoSAX/WFCs and RXTE/PCA X-ray instruments.
This catalog contains 136 bursts with recurrence times of less than
one hour, that come in multiples of up to four events, from 15 sources.
Short recurrence times are not observed from so-called ultra-compact
binaries, indicating that hydrogen burning processes play a crucial
role. As far as the neutron star spin frequency is known, these sources
all spin fast at over 500~Hz; the rotationally induced mixing may
explain burst recurrence times of the order of 10~min. Short recurrence
time bursts generally occur at all mass accretion rates where normal
bursts are observed, but for individual sources the short recurrence
times may be restricted to a smaller interval of accretion rate. The
fraction of such bursts is roughly 30\%. We also report the shortest
known recurrence time of 3.8~minutes.
\end{abstract}


\section{Introduction}

Type I X-ray bursts are thought to result from thermonuclear flashes
of hydrogen and/or helium in the envelope of neutron stars (\citealt{Woosley1976,Maraschi1977,Lamb1978}).
This material is accreted through Roche-lobe overflow from a lower-mass
companion star (low-mass X-ray binary, LMXB). Current one-dimensional
models successfully explain burst features such as the peak flux,
the fluence, decay time and recurrence time (e.g., \citealt{Woosley2004,Heger2007};
see \citealt{Wallace1981,Fujimoto1981,Fushiki1987ApJ} for earlier
work). During the flash, over $90\%$ of the accreted hydrogen and
helium is expected to burn to carbon and heavier elements (e.g., \citealt{Woosley2004}).
For the next flash to occur, a fresh layer of hydrogen/helium must
first be accreted. At typical accretion rates of up to approximately
$10^{-8}\,\mathrm{M_{\odot}yr^{-1}}$ this takes at least a few hours.

X-ray bursts have been observed since the 1970's (\citealt{Grindlay1976,1976Belian})
from approximately $90$ sources in our Galaxy, with recurrence times
of hours up to days (e.g., \citealt{Lewin1993,Strohmayer2006}). \citet{lewin76mnras}
reported the detection with \emph{SAS-3} of three bursts that were
separated by only $17$ and $4$~minutes. These bursts originated
from a crowded region and, therefore, source confusion cannot be ruled
out. In the 1980's similar recurrence times as short as $10$~minutes
were observed from both 4U~1608-522 with \emph{Hakucho} (\citealt{1608:murakami80pasj})
and from EXO~0748-676 with \emph{EXOSAT} (\citealt{0748:gottwald86apj,0748:gottwald87apj}).
This rare phenomenon implies that hydrogen and helium is left over
somewhere on the star after the initial flash, because the recurrence
time is too short to accrete enough fuel for the subsequent burst(s).
This is at odds with the current models, that predict an almost complete
burning of the available hydrogen and helium on the entire star surface.

\citet{Boirin2007} analyzed 158 hours of \emph{XMM-Newton} observations
of EXO~0748-676, which revealed short recurrence time bursts in groups
of two (doubles) and three (triples). This relatively large burst
sample revealed that on average bursts with a short recurrence time
($8$ to $20$~minutes) are less bright and energetic than bursts
with `normal' recurrence times (over $2$~hours). The fit of a black
body model to the burst spectrum shows a lower peak temperature, while
the emitting area is the same. The profiles of short recurrence time
bursts seemingly lack the long $50\,\mathrm{s}$ to $100\,\mathrm{s}$
tail caused by \textsl{rp}-process burning, which indicates that the
burst fuel contains less hydrogen. After a double or triple it takes
on average more time before another burst occurs, suggesting a more
complete burning of the available fuel.

\citet{Galloway2008catalog} showed that there are more sources that
show this behavior, with bursts occurring in groups of up to four
bursts, and with recurrence times as short $6.4$~minutes. The short
recurrence times were observed predominantly when the persistent flux
is between approximately $2\%$ and $4\%$ of the Eddington limited
flux. Furthermore, indications were found for the association of short
recurrence times and the accretion of hydrogen-rich material. The
shortest recurrence time previously reported is $5.4$~minutes (\citealt{Linares2009ATel}).

Different ideas have been put forward to explain this rare bursting
behavior. As most of the models only resolve the neutron star envelope
in the radial direction, it is possible that short recurrence time
bursts are due to multi-dimensional effects, such as the confinement
of accreted material on different parts of the surface, possibly as
the result of a magnetic field (e.g., \citealt{Melatos2005,Lamb2009}).
\citet{Boirin2007}, however, found that the different bursts originate
from an emitting area of similar size. Furthermore, the indication
of a different fuel composition for the bursts with short recurrence
times, argues against any scenario where accreted material of the
same composition burns on different parts of the surface.

The idea of a burning layer with an unburned layer on top has been
investigated (\citealt{1636:fujimoto87apj}). After the first layer
flashes, the second layer could be mixed down to the depth where a
thermonuclear runaway occurs. Mixing may be driven by rotational hydrodynamic
instabilities (\citealt{Fujimoto1988A&A}) or by instabilities due
to a rotationally induced magnetic field (\citealt{Piro2007,Keek2009}).
The mixing processes take place on the correct time scale of approximately
ten minutes. Although this scenario is able to explain many of the
observed aspects of short recurrence time bursts, it has not been
reproduced with a multi-zone stellar evolution code that includes
a full nuclear burning network. \citet{Taam1993} created models that
exhibit `erratic' bursting behavior, reminiscent of short recurrence
time bursts. Later versions of the employed code, however, no longer
produce this, most likely because of the inclusion of a more extensive
nuclear network (\citealt{Woosley2004}). As a different explanation
for the reignition, \citet{Boirin2007} suggested a waiting point
in the chain of nuclear reactions. There may be a point in the chain
where a decay reaction with a half life similar to the short recurrence
times stalls nuclear burning before continuing.

We use an improved version of the burst catalog compiled by \citet{Galloway2008catalog},
that is extended with the X-ray burst observations of the WFCs on-board
\emph{BeppoSAX} (e.g., \citealt{Cornelisse2003}). This is the largest
collection of X-ray bursts used in any study to date. This allows
us to study the short recurrence time phenomenon in much more detail.

\section{Observations and Analysis Methods}

\subsection{Nomenclature}

In this paper we name the different kinds of bursts using the conventions
of \citet{Boirin2007}. Bursts with recurrence (waiting) times shorter
than one hour are referred to as \emph{short waiting time (SWT)} bursts,
while longer recurrence time bursts are \emph{long waiting time (LWT)}
bursts. The one-hour boundary is chosen to discriminate between the
two distinct groups of bursts we observe (Sect.~\ref{sub:Recurrence-times}).
A burst \emph{event} is defined as a series of bursts, where any two
or more subsequent bursts are separated by a short waiting time. We
refer to an event with one, two, three, or four bursts as a \emph{single},
\emph{double}, \emph{triple}, or \emph{quadruple} burst, respectively.
Furthermore, we call a double, triple, or quadruple burst a \emph{multiple-burst
event}. For the bursts within a multiple-burst event, we use the terms
\emph{first burst }and \emph{follow-up burst}, where the former refers
to the LWT burst and the latter to any SWT burst in the event.

\subsection{Burst catalog}

Because SWT bursts are rare, we need a large sample of bursts to study
them well. We use a preliminary version of the Multi-INstrument Burst
ARchive (MINBAR), which is a collection of Type I bursts that are
observed with different X-ray instruments, and that are all analyzed
in a uniform way. MINBAR is a continuation of the effort which started
with the \emph{RXTE} PCA burst catalog by \citet{Galloway2008catalog},
combined with the many X-ray bursts observed with the WFCs on \emph{BeppoSAX}
(e.g., \citealt{Cornelisse2003}). The catalog will be presented in
full in a forthcoming paper. Currently it contains information on
3402 bursts from 65 sources, among which are 136 SWT bursts (MINBAR
version 0.4). In comparison, \citet{Galloway2008catalog} report 1187
bursts from 48 sources, among which are 84 bursts with a short recurrence
time, in that paper defined as $<30\,\mathrm{min}$ (MINBAR contains
110 SWT bursts using this criterion).

\subsection{Instruments\label{sub:Instruments}}

The \emph{Rossi X-ray Timing Explorer} (\emph{RXTE}) was launched
on December 30, 1995. One of the instruments on-board is the Proportional
Counter Array (PCA; \citealt{Jahoda2006}), consisting of five proportional
counter units (PCUs) which observe in the $1$ to $60\,\mathrm{keV}$
energy range. The PCA has a large collecting area of $8000\,\mathrm{cm^{2}}$
($1600\,\mathrm{cm^{2}}$ for each PCU). The PCUs are co-aligned and
have a collimator that gives a $1^{\circ}$ FWHM field of view. As
part of the primary mission objective of \emph{RXTE}, the PCA gathered
a large amount of exposure time on the galactic LMXB population. The
observations we use have an average duration of $78\,\mathrm{min}$.
We use all data that was publicly available in July 2008 (compared
to June 2007 for \citealt{Galloway2008catalog}).

The High-Energy X-ray Timing Experiment (HEXTE; \citealt{Gruber1996})
on \emph{RXTE} consists of two clusters of NaI/CsI scintillation detectors
that allow for observations in the $15$ to $250\,\mathrm{keV}$ energy
range. If available, we combine PCA and HEXTE observations to obtain
a broad band X-ray spectrum, when studying the persistent flux of
bursting sources.

The All-Sky Monitor (ASM) on \emph{RXTE} consists of three scanning
shadow cameras (SSCs) on a rotating beam that image a large part of
the sky each satellite orbit, in a series of short $90\,\mathrm{s}$
observations. The SSCs observe in the $1.5$ to $12\,\mathrm{keV}$
energy range.

A few months after \emph{RXTE}, the \emph{BeppoSAX} observatory was
launched in April 1996 (\citealt{1997Boella}), and it was operational
until April 30, 2002. On-board were two Wide-Field Camera's (WFCs)
which faced in opposite directions (\citealt{1997Jager}). The WFCs
were sensitive in the 2 to 28~keV band-pass and each camera imaged
at any time $40^{\circ}\times40^{\circ}$ of the sky using a coded
mask aperture. Bi-yearly observations of the Galactic Center were
performed, resulting in a large exposure time for many of the LMXBs
in the Galaxy (\citealt{2004ZandWFC}). the WFC observations have
a mean duration of $255\,\mathrm{min}$. We use all WFC data.

For the majority of the bursters, which are located near the Galactic
Center, we gather a total net exposure time of approximately $50$~days.

Both observatories were placed in a low Earth orbit of approximately
$96$ minutes. During the observation of a particular source, the
source is obscured by the Earth for a typical duration of approximately
$36$ minutes per orbit. Furthermore, when the satellites passed through
the South Atlantic Anomaly (SAA), the detectors were turned off to
prevent damage. This introduces data gaps of $13$ to $26$ minutes
long. The precise length of the different data gaps depends on the
latitude of the source with respect to the satellite orbit, which
was different for \emph{BeppoSAX} and \emph{RXTE}.

\subsection{Burst analysis}

We briefly discuss the process of the burst analysis. The generation
of data products for the different instruments is handled identically
to the studies by \citet{Galloway2008catalog} and \citet{Cornelisse2003}.
Our method of burst analysis is in principle the same as for those
studies, but extra care was taken to ensure that results from different
instruments are directly comparable. For more details of the analysis
of PCA and WFC data, we refer to \citet{Galloway2008catalog} and
\citet{Cornelisse2003}, respectively.

To find burst occurrences, we generate light curves for each instrument,
for each known bursting source. In the light curves we locate all
events that rise significantly above the background, e.g., exceeding
the mean flux level by at least four times the standard deviation.
These events are checked by eye for the characteristic profile of
a fast rise and an exponential-like decay. 

Next, time-resolved spectroscopy is performed on the candidate bursts.
We divide the individual bursts in time intervals, such that we obtain
for each interval a burst spectrum with similar and sufficient statistics.
The net burst spectrum is obtained by subtracting the spectrum of
the persistent emission observed during the entire observation, excluding
the burst (e.g., \citealt{2002Kuulkers}). We fit the burst spectra
with a black body model, taking into account the effects of interstellar
absorption (\citealt{1983Morrison}; solar abundances from \citealt{Anders1982}).
From this we find a black body temperature and radius, as well as
the unabsorbed flux. By extrapolating the fitted black body model
beyond the observed energy range, we obtain the bolometric unabsorbed
flux. Integrating the flux over the burst yields the burst fluence.

The decay time is determined by fits to the burst light curve. We
obtain the net burst light curve by subtracting the persistent flux,
as measured in the entire observation, excluding the burst. We start
fitting the decay when the flux drops below $90\%$ of the peak flux.
This has the advantage that we are less sensitive to the Poisson noise
at the peak or to the effects of radius expansion. The decay of many
bursts is fit well by two exponentials, with two exponential decay
times. For some bursts two exponentials did not yield a statistically
better fit than a single exponential. For those bursts we report only
a single decay time. It should be noted that not for all bursts a
good fit was obtained, but the `best' fit still provides a qualitative
description of the burst decay.

The determination of the burst decay time is currently not done uniformly
for the different instruments. For the PCA the bolometric flux is
used, while for the WFCs the flux in counts per second is used. We
compared the decay times obtained for $15$ bursts that have been
observed by both instruments. On average, the decay times for the
WFC bursts are $(22\pm7)\%$ longer than for the PCA bursts. Although
the decay time scales are not fully compatible across the different
instruments, we can still use them to differentiate between short
bursts and the longer bursts with the \textsl{rp}-process tail.

\subsection{Persistent emission and mass accretion rate\label{sub:Persistent-emission-and}}

We obtain the persistent flux for each burst by fitting the spectrum
from the entire observation, excluding the burst. By extending the
fitted spectral model (typically a comptonized spectrum or a black
body + power law; see \citealt{Galloway2008catalog} for details)
outside the observed energy range, we determine the bolometric correction.
The uncertainty in the bolometric correction can be as small as $10\%$,
when we combine \emph{RXTE} PCA and HEXTE observations to obtain a
broadband spectrum. The \emph{BeppoSAX} WFCs, however, do not have
such broad energy coverage, and some PCA observations suffer from
source confusion, which leads to an increased uncertainty in the bolometric
correction of up to $30\%$.

To convert flux to luminosity and fluence to energy, we multiply by
$4\pi d^{2}$, with $d$ the distance to the source. We use the distances
from \citet{Kuulkers2003} for globular cluster sources, and from
\citet{Galloway2008catalog} (distances from photospheric-radius expansion
bursts observed with the PCA) and \citet{Liu2007} for the other sources.
Most measurements of the distance have an uncertainty of the order
of $30\%$. Combined with a $30\%$ error in the bolometric correction,
this leads to an uncertainty in the luminosity and the fluence of
up to approximately $70\%$.

The persistent X-ray emission from an LMXB mainly originates from
the inner part of the accretion disk, and, therefore, is a measure
of the mass accretion rate (e.g, \citealt{1826:galloway04apj}). We
express the mass accretion rate $\dot{M}$ in terms of the Eddington
limited accretion rate $\mathrm{\dot{M}_{\mathrm{Edd}}}$ by equating
their ratio to the ratio of the persistent luminosity $L$ and the
Eddington luminosity for hydrogen accreting sources $\mathrm{L_{\mathrm{Edd}}}=2\cdot10^{38}\,\mathrm{erg\, s^{-1}}$:
$\dot{M}/\mathrm{\dot{M}_{\mathrm{Edd}}}=L/\mathrm{L_{\mathrm{Edd}}}$.
$\mathrm{L_{Edd}}$ depends on the neutron star mass and the hydrogen
fraction of the accreted material (e.g., \citealt{Bildsten1998}),
both of which are not known to great precision for most LMXBs. The
current observational constraints on the mass (\citealt{Lattimer2007})
introduce an uncertainty of several tens of percents. Furthermore,
we assume that the accretion process has an efficiency of $100\%$.
It is possible that part of the matter leaves the system in a jet,
such as observed in black-hole binaries (e.g., \citealt{Fender2005}).
In this paper, however, we neglect this possibility and assume the
luminosity is a good measure of the mass accretion rate. We also neglect
any anisotropy factors that may arise from the inclination of the
disk with respect to the line of sight; because the inclination is
ill-constrained for most LMXBs, we assume isotropic emission.

The combined uncertainties are quite large, but this only plays a
role when we compare different sources. It is, however, of no consequence
when we compare the bursts of any single source. We will still compare
different sources, but one must be careful to keep these uncertainties
in mind.

\section{Results}

\begin{table*}
\caption{\label{tab:Overview-of-16}Overview of 15 bursters with short recurrence
times.}

\begin{tabular}{lllllllll}
\hline 
Name & $\nu_{\mathrm{spin}}$ & $t_{\mathrm{exposure}}$ & MINBAR & Single & Double & Triple & Qua- & Remarks\tabularnewline
 & (Hz) & (days) & bursts &  &  &  & druple & \tabularnewline
\hline
\object{EXO~0748-676} & $552$$^{\mathrm{a}}$ & 83.1 & 269 & 251 & 9 & 0 &  & Triples: \citet{Boirin2007}; Appendix \tabularnewline
\object{GS~0836-429} &  & 30.9 & 17 & 17 & 0 &  &  & Double: \citet{aoki92pasj}\tabularnewline
\object{4U~1323-62} &  & 53.9 & 41 & 33 & 4 &  &  & \tabularnewline
\object{4U~1608-522} & $620$$^{\mathrm{b}}$ & 52.4 & 67 & 60 & 2 & 1 &  & \tabularnewline
\object{4U~1636-536} & $581$$^{\mathrm{c}}$ & 60.6 & 241 & 212 & 11 & 1 & 1 & \tabularnewline
\object{MXB~1658-298} & $567$$^{\mathrm{d}}$ & 49.8 & 27 & 25 & 1 &  &  & \tabularnewline
\object{4U~1705-44} &  & 60.1 & 125 & 108 & 5 & 1 & 1 & \tabularnewline
\object{XTE~J1710-281} &  & 47.9 & 18 & 16 & 1 &  &  & \tabularnewline
\object{4U~1735-44} &  & 54.7 & 50 & 38 & 6 &  &  & \tabularnewline
\object{2S~1742-294} &  & 56.3 & 269 & 240 & 10 & 3 &  & \tabularnewline
\object{EXO~1745-248} (Tz\,5)$^{\mathrm{f}}$ &  & 48.3 & 24 & 15 & 1 & 1 & 1 & Type II bursts? (\citealt{Galloway2008catalog})\tabularnewline
\object{4U~1746-37} (NGC~6641)$^{\mathrm{f}}$ &  & 55.1 & 31 & 15 & 8 &  &  & Two sources? (\citealt{Galloway2004AIPC})\tabularnewline
\object{SAX~J1747.0-2853} &  & 71.5 & 63 & 57 & 3 &  &  & \tabularnewline
\object{Aql~X-1} & $549$$^{\mathrm{e}}$ & 34.3 & 60 & 49 & 4 & 1 &  & \tabularnewline
\object{Cyg~X-2}$^{\mathrm{f}}$ &  & 55.7 & 55 & 45 & 5 &  &  & \tabularnewline
\hline
\end{tabular}

$^{\mathrm{a}}$ \citet{Galloway2009}

$^{\mathrm{b}}$ \citet{Muno2002}

$^{\mathrm{c}}$ \citet{Strohmayer1998}

$^{\mathrm{d}}$ \citet{Wijnands2001}

$^{\mathrm{e}}$ \citet{Zhang1998}

$^{\mathrm{f}}$ Source excluded from our analysis. See Sect.~\ref{sub:Source-selection}
for details.
\end{table*}

\subsection{Source selection\label{sub:Source-selection}}

The MINBAR catalog contains X-ray bursts from $65$ sources, $15$
of which exhibit SWT bursts (Table~\ref{tab:Overview-of-16}), i.e.,
bursts with a recurrence time shorter than one hour. We exclude the
Rapid Burster, which is known to exhibit Type II bursts, which are
not of thermonuclear origin (e.g., \citealt{Lewin1993}). Interestingly,
no multiple-burst events are detected from candidate and confirmed
ultra-compact binaries (UCXBs). The companion star in a UCXB is thought
to be an evolved star, donating hydrogen-poor matter to the neutron
star (\citealt{Zand2005}). For confirmed UCXBs the binary period
has been measured, while candidates are identified by tentative measurements
of the period, by a low optical to X-ray flux, or by stable mass transfer
at rates below $1\%$ of the Eddington limited rate. We employ the
list of (candidate) UCXBs from \citet{intZand2007}, which omits candidates
proposed on the basis of their X-ray spectrum, which is likely a less
reliable method. From these sources we find $229$ bursts, none of
which are SWT bursts. The frequent burster \object{4U~1728-34}
(GX~354-0) is suspected of being a UCXB based on its bursting behavior
(\citealt{Galloway2008catalog}). $543$ bursts of this source are
present in MINBAR, but no multiple-burst events. As this strongly
suggests that SWT bursts are limited to hydrogen-rich accretors, we
exclude in our studies the list of (candidate) UCXBs from \citet{intZand2007},
with the addition of 4U~1728-34.

GX~17+2 and Cyg~X-2 exhibit bursts at accretion rates close to the
Eddington limit (e.g., \citealt{2002Kuulkers}), while most LMXBs
do not show bursts above approximately $10\%$ of Eddington (e.g.,
\citealt{Paradijs1988,Cornelisse2003}). Of these two sources, Cyg~X-2
exhibits SWT bursts. Due to the high accretion rate, however, the
amount of matter accreted in between the bursts is enough to account
for the amount of fuel burned in the bursts. Furthermore, \citet{Galloway2008catalog}
find indications that (some of) the bursts of GX~17+2 and Cyg~X-2
could be Type II bursts. For these reasons we exclude these two sources
as well.

The bursts from EXO~1745-248, located in the globular cluster Terzan~5,
exhibit only weak evidence for cooling, which means that their thermonuclear
origin is not firmly established, and that they possibly are Type
II bursts. Another sign that the bursting behavior is anomalous, is
the fact that most sources in Table~\ref{tab:Overview-of-16} exhibit
more double bursts than triples and quadruples, while EXO~1745-248
has one of each. We exclude the source in our studies.

4U~1746-37, located in the globular cluster NGC~6641, exhibits both
faint and bright bursts. \citet{Galloway2004AIPC} found in PCA observations
that the faint bursts occur at very regular intervals, unaffected
by the occurrence of bright bursts, and vice versa. This lead to the
speculation that the faint bursts originate from a different LMXB
that is also located in NGC~6641. We, therefore, exclude the bursts
from this source.

After excluding the mentioned sources, we consider $44$ hydrogen-rich
accretors from which we observe $2274$ single, $56$ double, $7$
triple, and $2$ quadruple events.

\subsection{Source confusion}

While the WFCs are imaging instruments, the PCA is not. The PCA's
collimators restrict the field of view to $1^{\circ}$ FWHM. Especially
in crowded regions, such as near the Galactic Center, multiple X-ray
sources may be in the field of view. For the sources in Table~\ref{tab:Overview-of-16}
we check the \emph{RXTE} ASM light curves for any nearby bright X-ray
sources that were active at the time of PCA burst observations. This
is the case for 2S~1742-294 and SAX~J1747.0-2853. Those bursts,
for which we cannot reliably measure the persistent flux, are excluded
from our studies.

The problem is small for SAX~J1747.0-2853, because most of its bursts
were observed with the WFCs, and we have a reliable measurement of
the persistent flux for all three SWT bursts. For 2S~1742-294 the
problem of source confusion plays a role for $61$ bursts, including
$12$ out of $16$ SWT bursts. Most of these bursts, including all
SWT bursts, occur in the time interval MJD 52175--52195. The persistent
flux as measured with the WFCs for bursts from that period varies
by less than $10\%$. We assign the mean WFC persistent flux to those
PCA bursts.

\subsection{Data gaps and recurrence time\label{sub:Data-gaps-and}}

We define the recurrence time as the time since the previous burst
from the same source in the catalog. Due to the frequent data gaps
(Sect.~\ref{sub:Instruments}), we must keep in mind that these are
upper limits. Performing Monte Carlo simulations, we investigate the
effect of the data gaps on the number of observed SWT and LWT bursts.

We generate a series of $10^{6}$ burst occurrence times and check
which bursts fall in data gaps. We use EXO~0748-676 as a template:
we generate LWT and SWT bursts with an SWT fraction of $30\%$; we
use the mean LWT and SWT recurrence times $3.0$~hours and $12.7$~minutes,
respectively (\citealt{Boirin2007}); we position the bursts in time
following a Gaussian distribution around the LWT or SWT $t_{\mathrm{recur}}$,
with a width of $16\%$ of either $t_{\mathrm{recur}}$ (mean variability
in persistent flux of a series of persistent sources; \citealt{kee06})
to model variations in the mass accretion rate. To check which of
these bursts would be observed in the presence of data gaps, we assume
a $96$~min. satellite orbit, containing a $36$~min. data gap due
to Earth occultation. The presence and duration of data gaps due to
the South-Atlantic Anomaly (SAA) depends on the position of the source
and the satellite at the time of the observation. We model SAA data
gaps by placing a $20$~min. gap at a random phase of each orbit.

\begin{figure}
\includegraphics[clip]{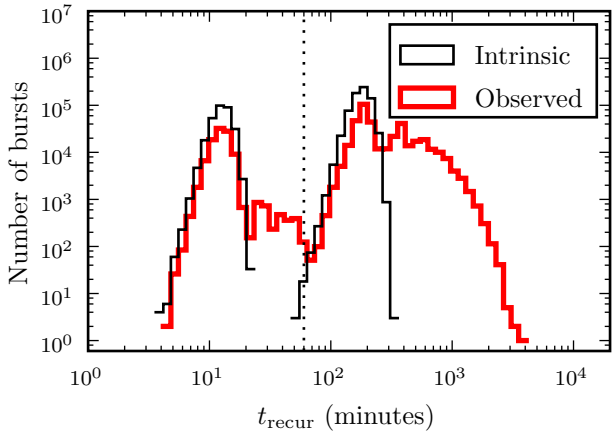}

\caption{\label{fig:data gaps}Histogram of simulated recurrence times $t_{\mathrm{recur}}$.
The \emph{intrinsic} distribution is from a Monte Carlo simulation
of $10^{6}$ bursts with SWT and LWT recurrence times as well as SWT
fraction from EXO~0748-676 (\citealt{Boirin2007}). The \emph{observed}
distribution includes the effect of data gaps due to Earth occultation
and the South-Atlantic Anomaly. All bursts with $t_{\mathrm{recur}}<60\,$min.
(dotted line) are considered SWT bursts.}

\end{figure}

Only $49\%$ of the generated bursts is `observed', the rest coincide
with a data gap. When a burst is missed, the recurrence time of the
next burst, as seen from the previously detected burst, is incorrectly
found to be longer: in the distribution of observed recurrence times
a long tail of bursts is present at longer $t_{\mathrm{recur}}$ than
present in the original distribution (Fig.~\ref{fig:data gaps}).
There are also bursts in between the Gaussian peaks of SWT and LWT
bursts, but their number is less than $10\%$ of the total number
of SWT bursts. Because most of these bursts have $t_{\mathrm{recur}}<60$~min.,
we still consider them SWT bursts.

\begin{figure}
\includegraphics[clip]{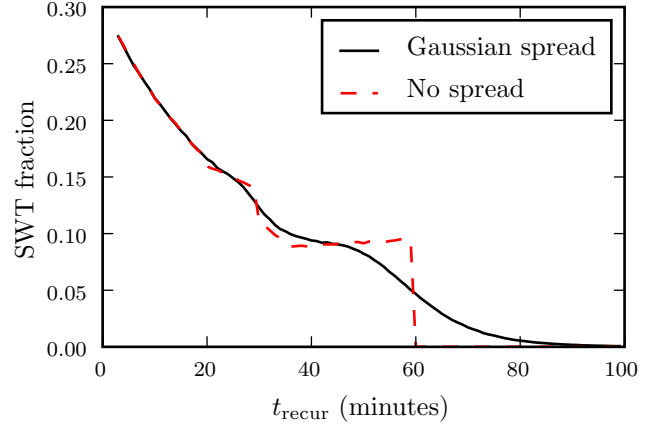}

\caption{\label{fig:swtfrac_data gaps}SWT fraction as a function of the SWT
recurrence time $t_{\mathrm{recur}}$. The solid line allows for variation
of $t_{\mathrm{recur}}$ following a Gaussian distribution, while
the dashed line does not. The latter drops to $0$ at $t_{\mathrm{recur}}=60\,\mathrm{min}.$,
which is the largest SWT $t_{\mathrm{recur}}$ we consider.}

\end{figure}

The fraction of bursts that are SWT bursts, is reduced when one or
more bursts in a multiple-burst event occur during a data gap. While
we start our simulations with an SWT fraction of $30\%$, the observed
distribution has a fraction of $20\%$. We repeat the simulations
with different values of the SWT $t_{\mathrm{recur}}$. The total
fraction of detected bursts remains the same, but the SWT fraction
drops from $30\%$ to $10\%$ for increasing $t_{\mathrm{recur}}$,
until $t_{\mathrm{recur}}$ exceeds the duration of the Earth occultation
data gap (Fig.~\ref{fig:swtfrac_data gaps}). %

\begin{figure}
\includegraphics[clip]{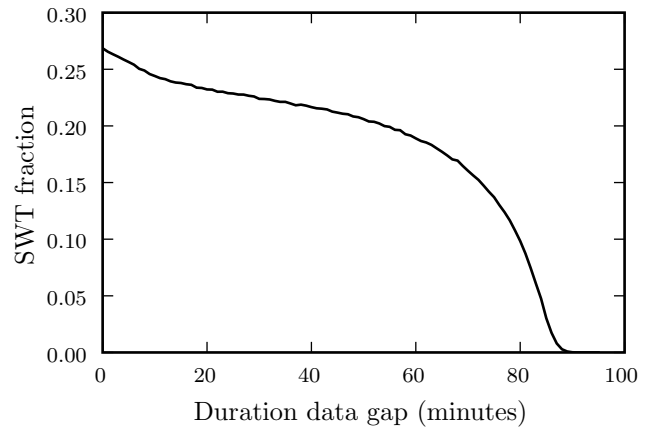}

\caption{\label{fig:swtfrac_data gaps-1}SWT fraction as a function of the
duration of the Earth-occulation data gaps. The SWT fraction drops
to $0$ when the data gaps approach the $96$~min. duration of the
satellite orbit.}

\end{figure}

Repeating the simulations with different durations of the Earth-occultation
data gap, the SWT fraction drops by only a few percent for longer
data gaps, until the fraction quickly goes to $0$, when an SWT recurrence
time no longer fits in the observed part of the orbit (Fig.~\ref{fig:swtfrac_data gaps-1}).

A similar decrease of the SWT fraction is expected if the duration
of the observation is less than the SWT recurrence time. While the
average exposure times of both the PCA and the WFCs exceed one hour,
a substantial part of the observations had a shorter recurrence time:
$19\%$ of the WFC exposures and $66\%$ of the PCA exposures were
shorter than one hour.

\subsection{Detection limits}

\begin{figure}
\includegraphics[clip]{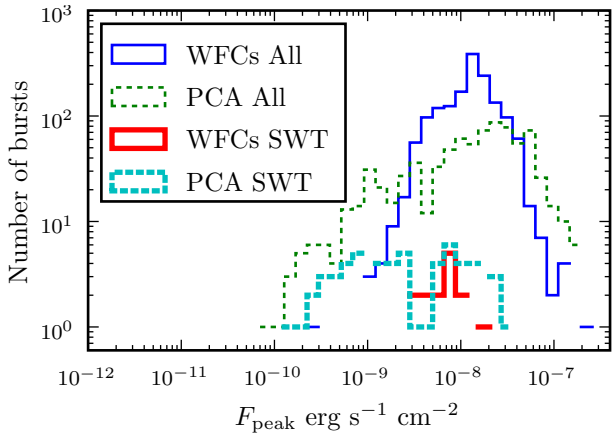}

\caption{\label{fig:instruments}Histogram of observed peak burst flux $F_{\mathrm{peak}}$
for the WFCs and the PCA. Additionally we show the distributions for
the short recurrence time bursts.}

\end{figure}
The instruments we use have different detection limits. The WFCs had
a substantially larger field of view than the PCA, which results in
a higher background level. Furthermore, the PCA has a $46$ times
larger collecting area than each WFC. Consequently, we are able to
find fainter bursts in PCA data than in WFC data (Fig.~\ref{fig:instruments}).
This is especially important for the SWT bursts, as they have been
found to be on average fainter than the LWT bursts (e.g., \citealt{Boirin2007}).
In the PCA data we find SWT bursts with peak flux $F_{\mathrm{peak}}$
as low as $1.6\cdot10^{-10}\,\mathrm{erg\, cm^{-2}\, s^{-1}}$, while
in the WFC data the faintest SWT burst has $F_{\mathrm{peak}}=3.6\cdot10^{-9}\,\mathrm{erg\, cm^{-2}\, s^{-1}}$.
As a result, we find many more SWT bursts with the PCA: for the PCA
we find 76 SWT bursts out of a total of 910 bursts, and for the WFCs
we find 14 SWT bursts out of 1560 bursts.

\subsection{\label{sub:Recurrence-times}Recurrence times}

\begin{figure*}
\includegraphics[clip]{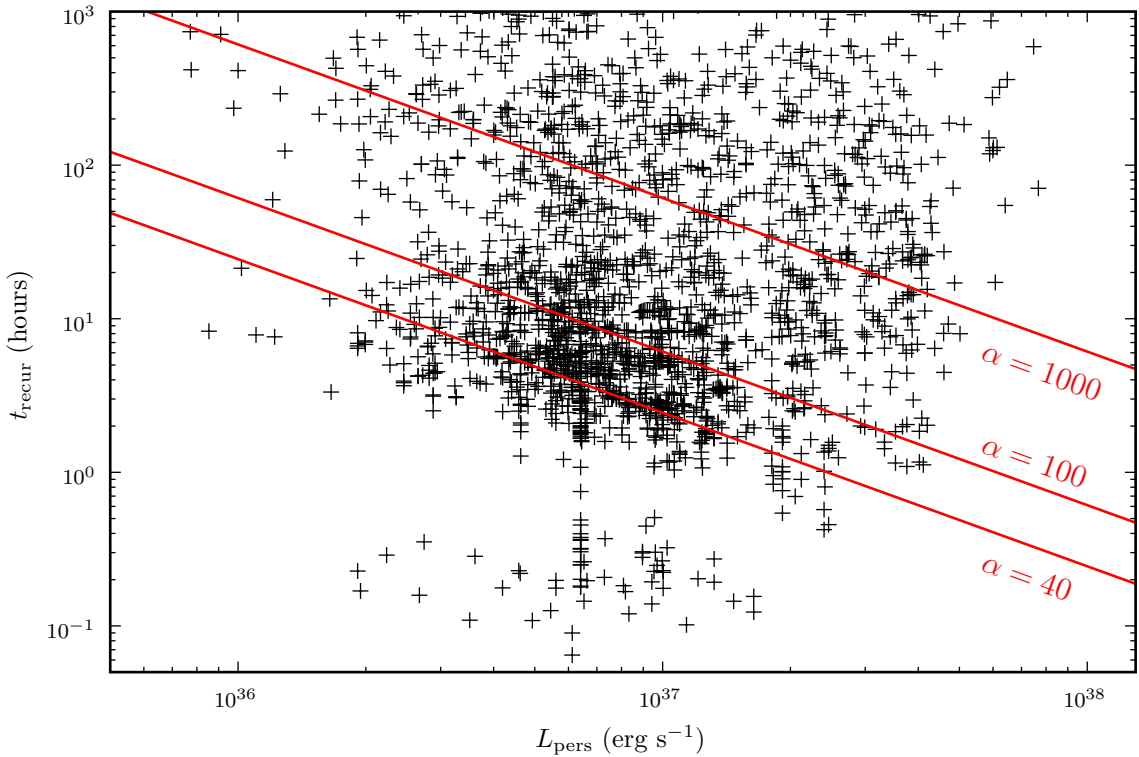}

\caption{\label{fig:twait_raw}Observed recurrence time $t_{\mathrm{recur}}$
as a function of the persistent luminosity $L_{\mathrm{pers}}$ for
$2415$ bursts from $44$ LMXBs. Due to the presence of data gaps,
values of $t_{\mathrm{recur}}$ exceeding one hour are upper limits
to the real recurrence time. We show bursts with $t_{\mathrm{recur}}<10^{3}\,\mathrm{hour}$.
The lines represent constant values of the average $\alpha$-parameter
as indicated (see text). Note that the crosses indicate only the positions
of the data points, not the uncertainties. The shortest recurrence
time is $3.8$~minutes.}

\end{figure*}

We plot for all bursts the recurrence time $t_{\mathrm{recur}}$ as
a function of the persistent luminosity $L_{\mathrm{pers}}$ (Fig.~\ref{fig:twait_raw}).
While most bursts have a recurrence time of at least several hours,
there is also a group of SWT bursts with $t_{\mathrm{recur}}<1$~hour.
There is an intrinsic spread in $t_{\mathrm{recur}}$ due to, for
example, variations in the mass accretion rate, or variations in the
temperature in the neutron star envelope. We investigate whether the
short recurrence times can be explained as the tail of the distribution
of the long recurrence times. For this distribution we assume a Gaussian,
even though it is not certain whether this is correct far from the
mean. The data gaps modify the observed distribution, especially towards
longer $t_{\mathrm{recur}}$ (Sect.~\ref{sub:Data-gaps-and}). The
leading part of the Gaussian, however, is not modified substantially,
apart from the overall lower number of observed bursts due to the
lower net exposure time (Fig.~\ref{fig:data gaps}). We fit a Gaussian
to the distribution of recurrence times between $1$ and $3$~hours
with the center fixed at $3$~hours. Extrapolating the best fit towards
shorter $t_{\mathrm{recur}}$, we predict $5.6$ SWT bursts. The Poisson
probability for the observed number of SWT bursts of $76$, is negligibly
small ($P\lesssim10^{-10}$). Therefore, the short recurrence times
follow a separate distribution.

There is a separation between bursts with short ($\lesssim0.5$~hour)
and long ($\gtrsim1$~hour) recurrence times. Above $L_{\mathrm{pers}}\gtrsim6\cdot10^{36}\,\mathrm{erg\, s^{-1}}$,
however, there are some bursts that have recurrence times of $30$
to $60$~minutes. Comparing the recurrence time distributions of
SWT bursts ($t_{\mathrm{recur}}<1$~hour) at persistent luminosities
below and above $L_{\mathrm{pers}}=6\cdot10^{36}\,\mathrm{erg\, s^{-1}}$,
a KS-test yields $P=0.16$, which means we can exclude at $84\%$
that both distributions are the same. This is not a strong constraint
constraint, mainly due to the small number of bursts with $L_{\mathrm{pers}}<6\cdot10^{36}\,\mathrm{erg\, s^{-1}}$.
It is, however, consistent with the behavior of EXO~0748-676 during
EXOSAT, XMM-Newton, and Chandra observations, that resulted in relatively
large data sets, where SWT bursts with recurrence times exceeding
$30$~minutes only occurred when the persistent flux was larger (\citealt{Gold1968,Gottwald1987,Boirin2007};
Appendix).

There are SWT bursts at all values of $L_{\mathrm{pers}}$ where LWT
bursts are observed, with the possible exception of $L_{\mathrm{pers}}\gtrsim3\cdot10^{37}\,\mathrm{erg\, s^{-1}}$,
although this may be a statistical effect due to the lower number
of bursts.

The $\alpha$-parameter is defined as the ratio of the persistent
fluence between bursts and the burst fluence. Assuming a burst fluence
of $2.2\cdot10^{39}\,\mathrm{erg}$ --- the average fluence of MINBAR
single bursts from hydrogen-rich accretors --- we draw lines of constant
$\alpha$; $\alpha\simeq40$ is a typical value for many bursters,
while $\alpha\simeq1000$ is observed for superbursters (e.g., \citealt{Zand2003}).
The effective $\alpha$-value for the SWT bursts is far below $\alpha=40$,
which is the expected value for thermonuclear ignition of mixed hydrogen/helium
fuel, highlighting the requirement for ignition of unburned fuel left-over
from the previous burst.

We find the shortest recurrence time reported so-far%
\footnote{\citet{1658:wijnands02apj} found a pair of candidate bursts from
MXB~1659-298 only $\sim50\,\mathrm{s}$ apart. The first flare, however,
is relatively weak and lacks the cooling characteristic for Type I
bursts. The second flare has been identified as $\gamma$-ray burst
GRB 990419C (HEASARC IPNGRB catalog).%
}: $3.8$~minutes for a double burst from 4U~1705-44, detected with
the WFCs at MJD~51233.89.

\subsection{Accretion rate dependence\label{sub:Accretion-rate-dependence}}

\begin{figure*}
\includegraphics[clip]{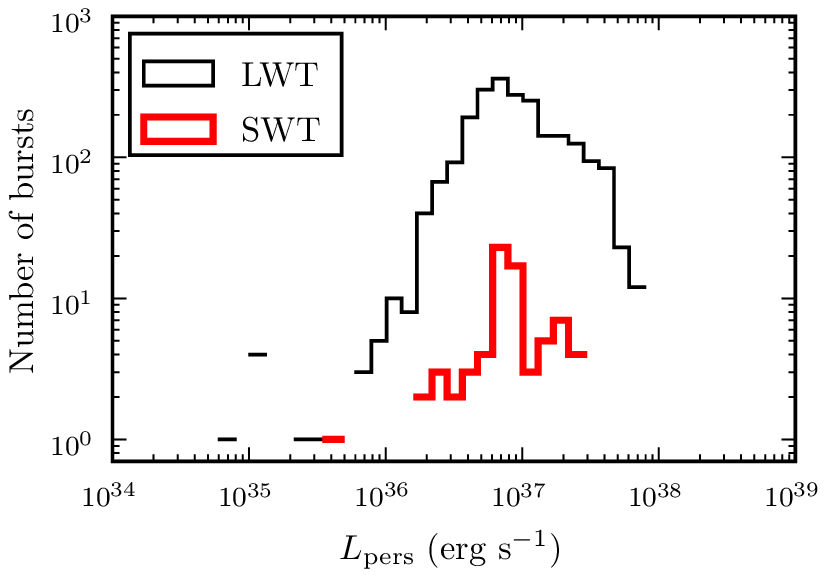}\includegraphics[clip]{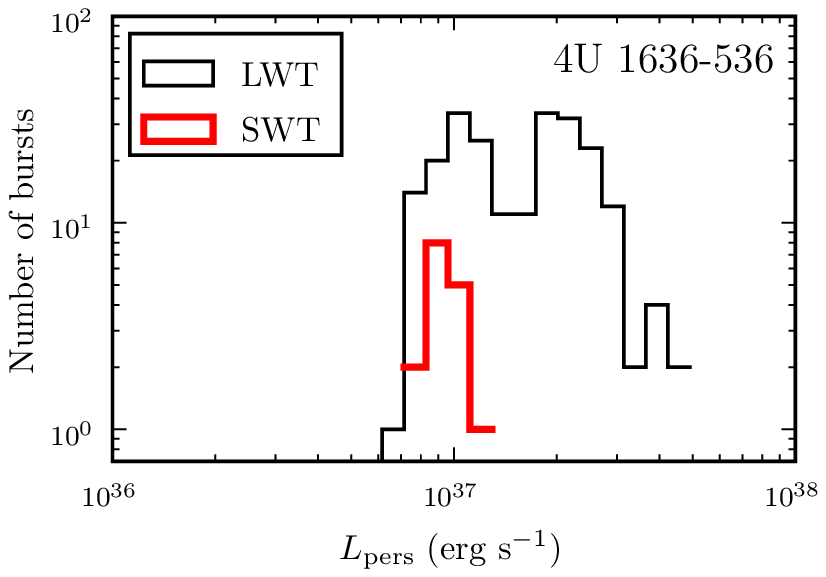}

\includegraphics{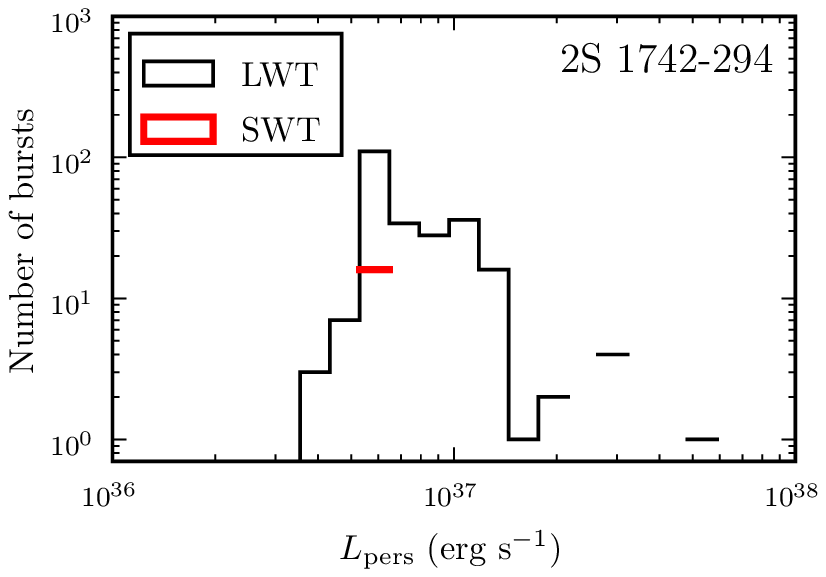}\includegraphics{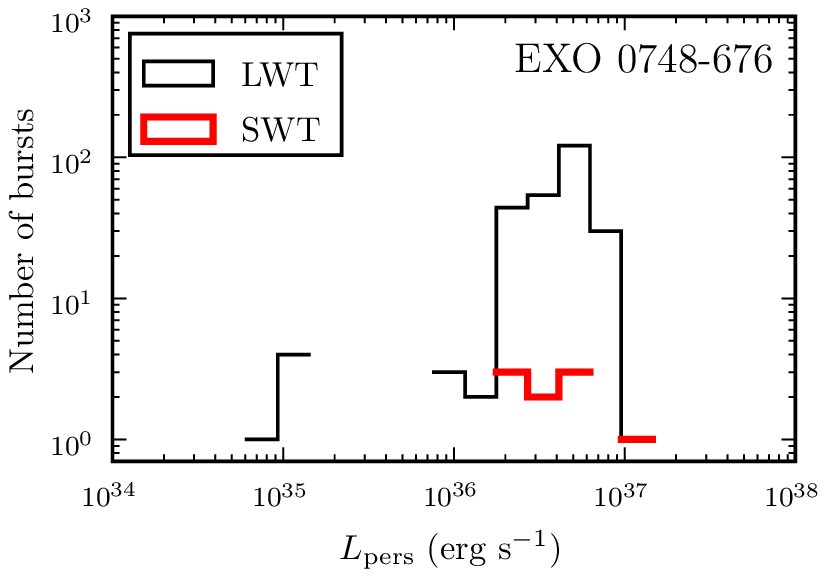}

\includegraphics{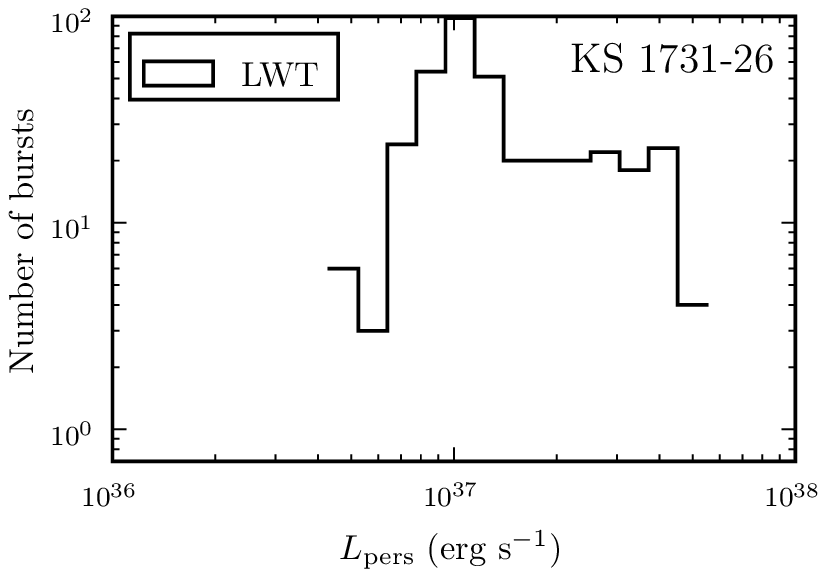}\includegraphics{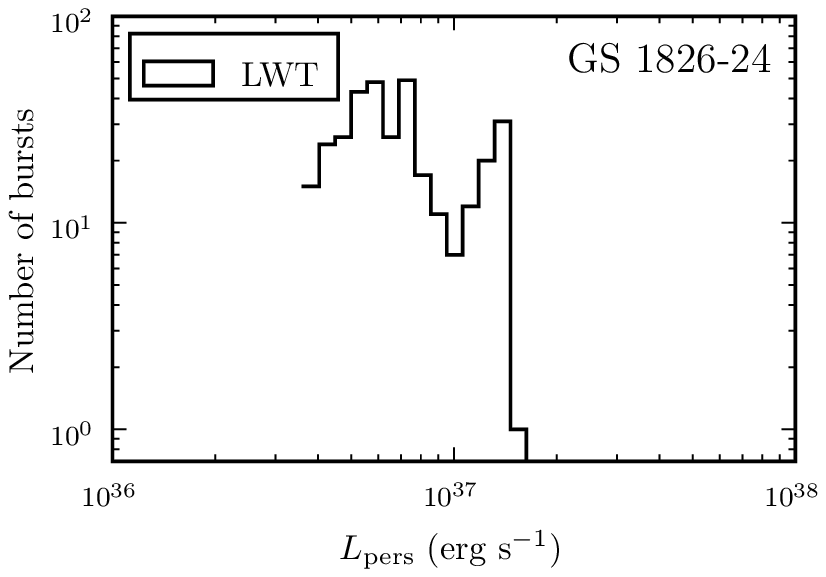}

\caption{\label{fig:fpers-histo}Histogram of bolometric persistent luminosity
$L_{\mathrm{pers}}$ of LWT and SWT bursts, for all hydrogen-rich
accretors (top-left) and for five individual sources. There is a dip
in the distribution for all sources around $L_{\mathrm{pers}}\simeq1.7\cdot10^{37}\,\mathrm{erg\, s^{-1}}$.
This is due to similar dips in the distributions of 4U~1636-53 and
2S~1742-294, as well as peaks in the distributions of KS~1731-26
and GS~1826-24 at slightly lower $L_{\mathrm{pers}}$.}

\end{figure*}

We use the persistent luminosity $L_{\mathrm{pers}}$ as a measure
of the mass accretion rate (Sect.~\ref{sub:Persistent-emission-and}).
A Kolmogorov-Smirnov (KS) test finds that the distributions of all
SWT and LWT as a function of $L_{\mathrm{pers}}$ are compatible ($P=0.10$;
Fig.~\ref{fig:fpers-histo}): there are multiple-burst events at
all mass accretion rates where normal bursts occur. This does not
hold, however, for all individual sources. We investigate a few frequent
bursters more closely. For EXO~0748-676 the distributions of single
and multiple bursts are compatible ($P=0.73$); 4U~1636-53 and 2S~1742-294
exhibit multiple bursts only in a small $L_{\mathrm{pers}}$ interval,
while single bursts occur in a wider range of $L_{\mathrm{pers}}$.
The small $L_{\mathrm{pers}}$ intervals for these two sources do
not seem to coincide. The uncertainty in the luminosity, however,
is large (Sect.~\ref{sub:Persistent-emission-and}), so the intervals
may still be consistent.

We find SWT bursts for only $15$ out of $44$ hydrogen-rich accretors.
For most of the sources we can attribute the lack of SWT bursts to
the low number of bursts detected per source. There are, however,
two bursters, \object{KS~1731-260} and \object{GS~1826-24},
from which we have detected over $300$ LWT bursts per source, but
no SWT bursts.

The position in so-called color-color diagrams, $S_{\mathrm{Z}}$,
is regarded as a tracer of the mass accretion rate (\citealt{1989Hasinger}).
We compared the distribution of $S_{\mathrm{Z}}$ for LWT and SWT
bursts observed with the PCA from eight sources for which $S_{\mathrm{Z}}$
is well defined: 4U~1608-522, 4U~1636-536, 4U~1702-429, 4U~1705-44,
4U~1728-34, KS~1731-260, Aql~X-1, and XTE~J2123-058, omitting
4U~1746-37 as explained in Sect.~\ref{sub:Source-selection} (\citealt{Galloway2008catalog}).
While LWT bursts have associated $S_{\mathrm{Z}}$ values of up to
$2.8$, SWT bursts all have $S_{\mathrm{Z}}\lesssim2$. Therefore,
we find that SWT bursts are restricted to the so-called island state,
while LWT bursts also occur in the `banana' branch. A KS-test yields
$P\simeq10^{-3}$, confirming that the $S_{\mathrm{Z}}$ distributions
for LWT and SWT bursts are different. Most of the SWT bursts from
the frequent burster 4U~1636-53 occurred with $1.5\lesssim S_{\mathrm{Z}}\lesssim2.0$.
KS~1731-260 exhibits only a few LWT bursts in that range. This suggests
that SWT bursts occur mainly in a small $S_{\mathrm{Z}}$ interval,
and that the lack of SWT bursts from the latter source is caused by
the low number of observed bursts in that range. Note that we observe
SWT bursts with $S_{\mathrm{Z}}$ as low as $0.8$, so the interval
is not the same for all sources. %

\subsection{Frequency of short vs. long recurrence times\label{sub:Frequency-of-short}}

\begin{figure*}
\includegraphics[clip]{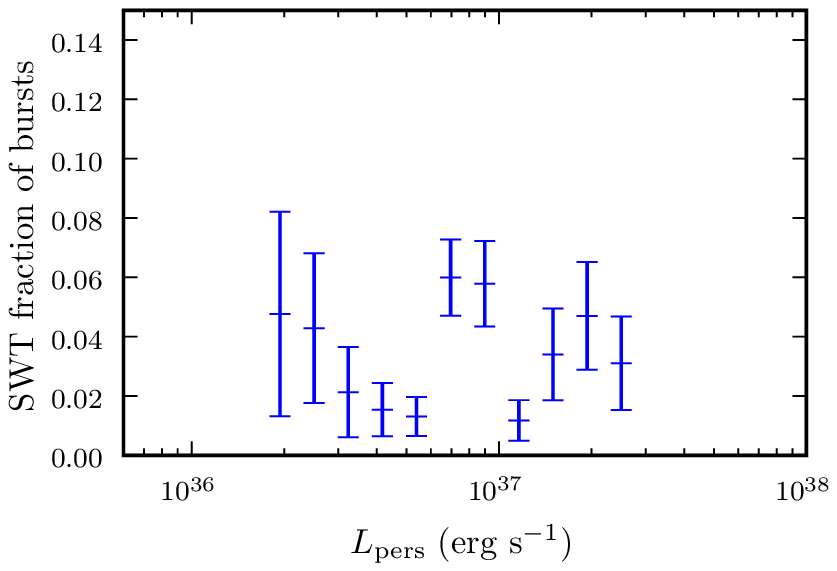}\includegraphics[clip]{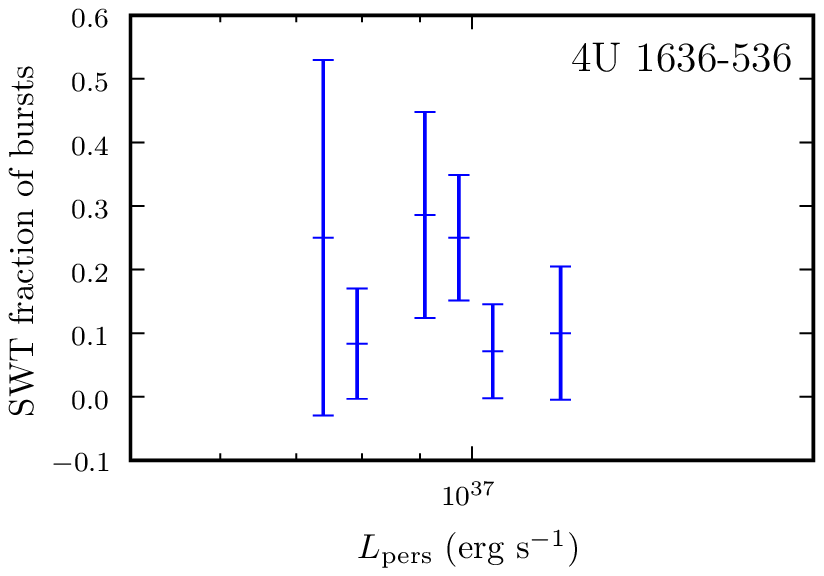}

\includegraphics{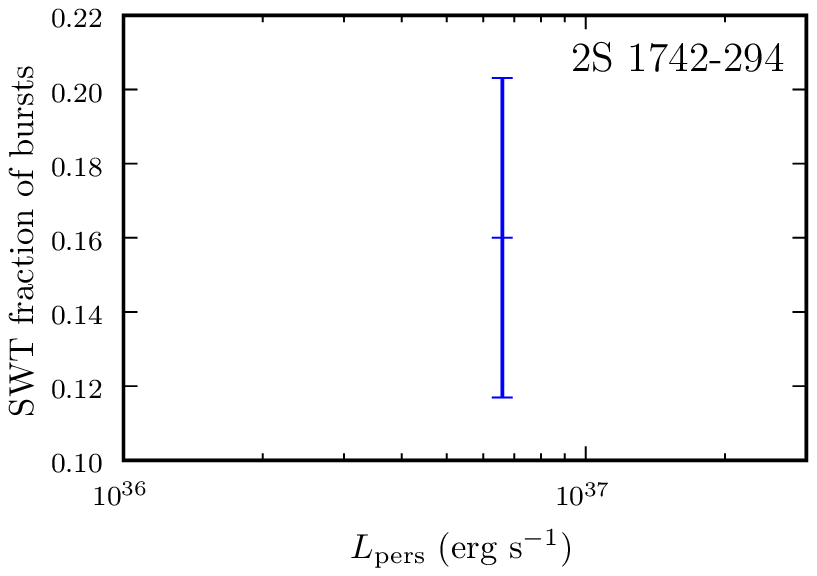}\includegraphics{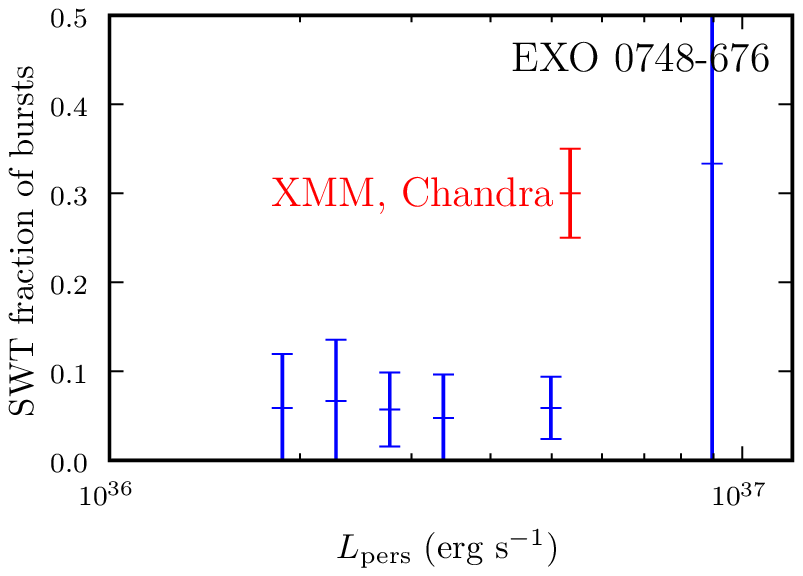}

\caption{\label{fig:multiplefrac}The fraction of bursts that have a short
recurrence time as a function of the persistent luminosity $L_{\mathrm{pers}}$,
for all hydrogen-rich accretors (top-left) and for three individual
sources. For each plot $L_{\mathrm{pers}}$ is divided into $30$
logarithmically spaced bins, and for each bin the SWT fraction is
calculated from the bursts in the corresponding plots in Fig.~\ref{fig:fpers-histo}.
For EXO~0748-676 one SWT burst is observed at higher $L_{\mathrm{pers}}$,
which results in a data point with large uncertainty.}

\end{figure*}
We consider $2415$ bursts from hydrogen-accreting sources. $76$
have a short recurrence time: the overall SWT fraction is $(3.1\pm0.4)\%$.
The 1-$\sigma$ uncertainty is derived from the Poisson uncertainties
in the number of (SWT) bursts. As a function of persistent luminosity
there is some variation in the SWT fraction, but there is no clear
trend (Fig.~\ref{fig:multiplefrac}). The weighted mean of the SWT
fraction in the range of $L_{\mathrm{pers}}$ where SWT bursts are
observed is $(2.3\pm0.3)\%$ for all hydrogen-rich accretors, $(13\pm4)\%$
for 4U~1636-536, $(16\pm4)\%$ for 2S~1742-294 and $(6\pm2)\%$
for EXO~0748-676. 

We compare these fractions to \emph{XMM-Newton} and \emph{Chandra}
observations of EXO~0748-676. In 2003 the \emph{XMM-Newton} EPIC
PN observed double and triple bursts from EXO~0748-676 with an SWT
fraction of $(32\pm7)\%$ (\citealt{Boirin2007}). \citet{Homan2003}
find in \emph{XMM-Newton} EPIC PN and MOS observations from 2000 and
2001 $4$ single and $4$ double bursts, which results in an SWT fraction
of $(33\pm19)\%$. \emph{Chandra} ACIS-S observations from 2001 and
2003 exhibit $41$ bursts with an SWT fraction of $(27\pm9)\%$ (see
Appendix). The weighted mean of these rates is $(30\pm5)\%$. An important
difference in the data of on the one hand \emph{XMM-Newton} and \emph{Chandra},
and on the other hand \emph{RXTE} and \emph{BeppoSAX}, is the frequent
data gaps in observations of the latter observatories. From Monte
Carlo simulations we found that data gaps reduce an SWT fraction of
$30\%$ to $20\%$ (Sect.~\ref{sub:Data-gaps-and}). This is significantly
higher than the $(6\pm2)\%$ we obtain from MINBAR. We repeat the
simulations for 4U~1636-536 and 2S~1742-294, using the mean SWT
recurrence times from MINBAR: $16.2$~min. and $20.5$~min., respectively.
The obtained SWT fractions, respectively $18\%$ and $16\%$, are
consistent within $1.3\sigma$ with the fractions from MINBAR.

The discrepancy for EXO~0748-676 may arise because of the fact that
the WFCs are less sensitive to fainter bursts than the PCA or the
instruments on \emph{XMM-Newton} and \emph{Chandra}. Taking only the
PCA bursts into account, we find SWT fractions of $(11\pm6)\%$ for
EXO~0748-676, $(13\pm4)\%$ for 4U~1636-53, and $(23\pm6)\%$ for
2S~1742-294, all of which are within $1.5\sigma$ from the fractions
found from the Monte Carlo simulations. 

Two frequent bursters exhibit no SWT bursts: KS~1731-26 and GS~1826-24.
Combining the number of LWT bursts observed from these sources, we
derive an upper limit to the SWT fraction of $0.14\%$. This is over
$20$ times smaller than the SWT fraction we find for all hydrogen-accreting
sources combined.

\subsection{Temperature and energetics}

\begin{figure}
\includegraphics[clip]{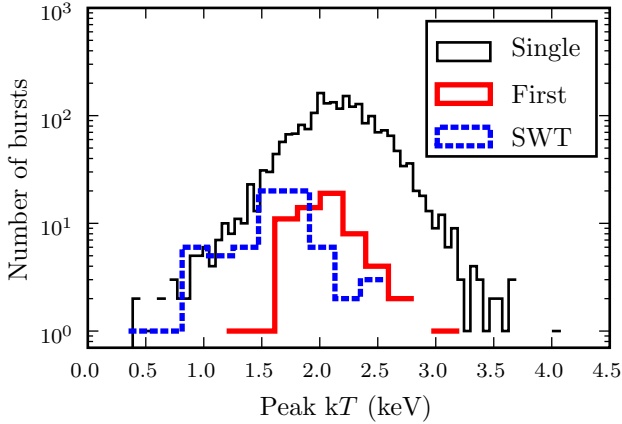}

\caption{\label{fig:peak temp}Histogram of peak black-body temperature of
single bursts, the first bursts in multiple events, and the SWT bursts.}

\end{figure}

From time-resolved spectral analysis of the bursts, we obtain the
black-body temperature and the burst energetics. The peak temperature
of the first bursts in multiple-burst events is on average higher
than the peak temperature of the SWT bursts (Fig.~\ref{fig:peak temp}).
A KS test shows that the temperature distributions for single bursts
and first bursts are not compatible ($P\simeq10^{-12}$).

\begin{figure}
\includegraphics[clip]{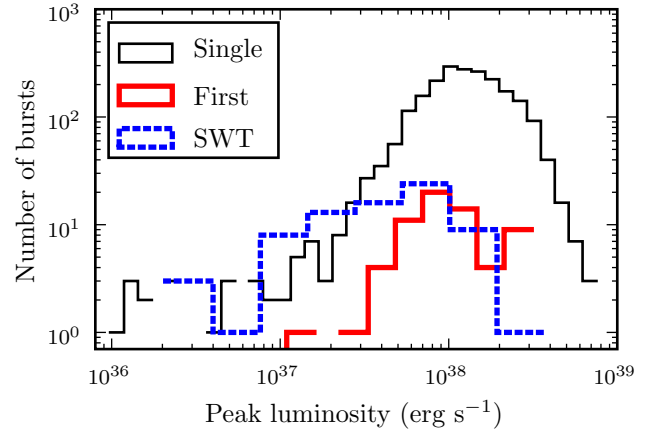}

\caption{\label{fig:peak luminosity}Histogram of bolometric peak luminosity
of single bursts, the first bursts in multiple events, and the SWT
bursts.}

\end{figure}

\begin{figure}
\includegraphics[clip]{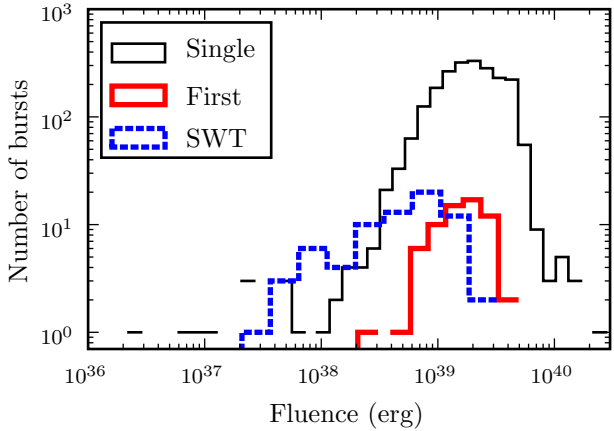}

\caption{\label{fig:fluence}Histogram of bolometric fluence of single bursts,
the first bursts in multiple events, and the SWT bursts.}

\end{figure}

\begin{figure}
\includegraphics[clip]{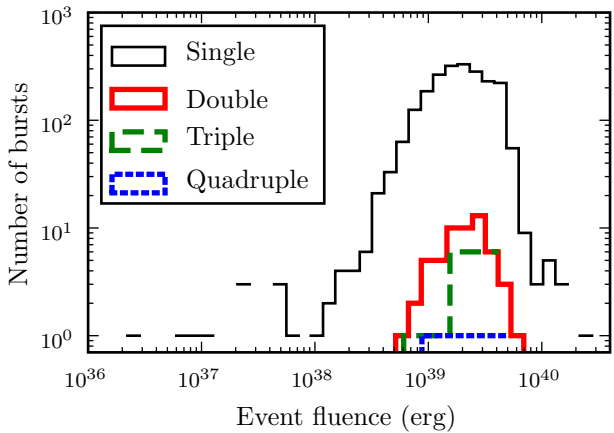}

\caption{\label{fig:event fluence}Histogram of the summed bolometric fluence
of all bursts in the different multiple-burst events.}

\end{figure}

We compare the peak luminosity and the fluence of single bursts to
those in multiple-burst events (Fig.~\ref{fig:peak luminosity},
Fig.~\ref{fig:fluence}). The SWT bursts have on average a lower
peak luminosity and a lower fluence. The energetics of the first bursts
does not follow the same distributions as the single bursts ($P\lesssim10^{-2}$).
By adding the fluence of all bursts in a multiple-burst event, we
calculate the total event fluence (Fig.~\ref{fig:event fluence}).
Multiple-burst events are on average more energetic than single burst
events, but do not have a higher fluence than the most energetic single
bursts.

Summarizing, SWT bursts are on average weaker and cooler than LWT
bursts, but the combined fluence in multiple-burst events is on average
$8\%$ higher than the fluence of single-burst events.

\subsection{Decay time scale}

\begin{figure}
\includegraphics[clip]{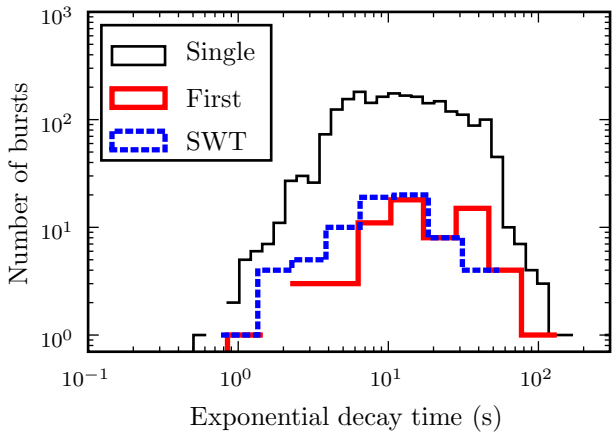}

\caption{\label{fig:decay time}Histogram of the exponential decay time for
single bursts of hydrogen-rich accretors, as well as the first and
SWT bursts in multiple events.}

\end{figure}

The two component exponential provides a good fit to many bursts.
If the two-component decay does not provide a significantly better
fit than the one-component, we use the latter. We compare the longest
decay time scales of all bursts in Fig.~\ref{fig:decay time}. On
average, the SWT bursts decay faster than the other bursts. KS tests
show that all distributions are incompatible ($P\lesssim10^{-2}$).

\subsection{Neutron stars with SWT bursts spin fast}

\begin{figure}
\includegraphics[clip]{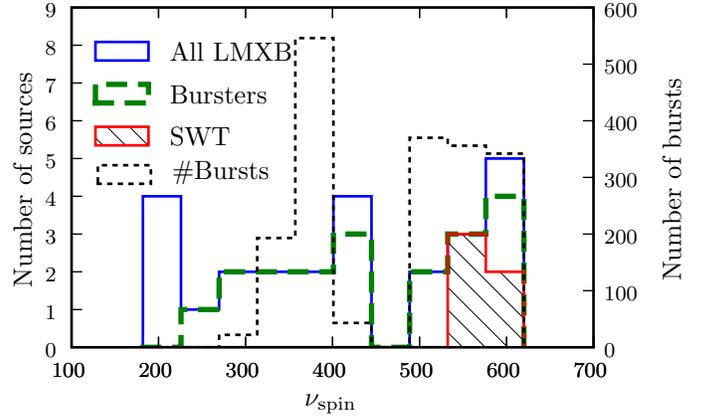}

\caption{\label{fig:ns spin}Histogram of the neutron star spin $\nu_{\mathrm{spin}}$
determined from X-ray observations, for all LMXBs with known spin,
for the known bursters and for the sources that exhibit multiple-bursts
(Table~\ref{tab:Overview-of-16}). The latter are concentrated at
the high end of the distribution of known spins. The dotted line indicates
the number of bursts in the MINBAR catalog for the bursters in each
bin (right-hand axis).}

\end{figure}

The spin frequency $\nu_{\mathrm{spin}}$ of an accreting neutron
star is measured in observations of accretion-powered pulsations or
burst oscillations. Both mechanisms are thought to arise from hotter
and, hence, brighter spots on the surface rotating in and out of view.
Currently $\nu_{\mathrm{spin}}$ is known for $25$ accreting neutron
stars, $19$ of which are bursters (e.g., \citealt{Galloway2008}).
Among them are $5$ sources that exhibit SWT bursts (Table~\ref{tab:Overview-of-16}).
They are concentrated towards the high-frequency part of the distribution
for all bursting sources (Fig.~\ref{fig:ns spin}). Since we only
have five multiple-bursting sources with a known spin frequency, we
cannot exclude that this bias is the result of the small sample. 

There could be a selection effect: sources with a higher mass accretion
rate accrete more angular momentum, causing them to spin up faster.
Furthermore, their burst rate is higher, making it easier to detect
a rare multiple-burst event. We check the MINBAR catalog for the number
of bursts from these sources as a function of the spin frequency (Fig.~\ref{fig:ns spin}).
There is roughly an equal number of bursts observed at higher and
at lower $\nu_{\mathrm{spin}}$, so the selection effect is not present.

There are a few hydrogen-accreting sources of which we detected a
large number of bursts, but no SWT bursts (Sect.~\ref{sub:Accretion-rate-dependence}).
One of these sources, KS~1731-260, is known to have a large spin
frequency of $\nu_{\mathrm{spin}}=524\,\mathrm{Hz}$ (\citealt{Smith1997}).

\subsection{Quadruple burst from 4U~1636-53}

\begin{figure}
\includegraphics[clip]{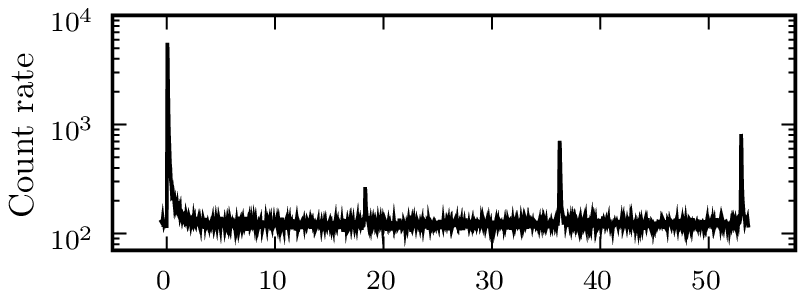}

\includegraphics[clip]{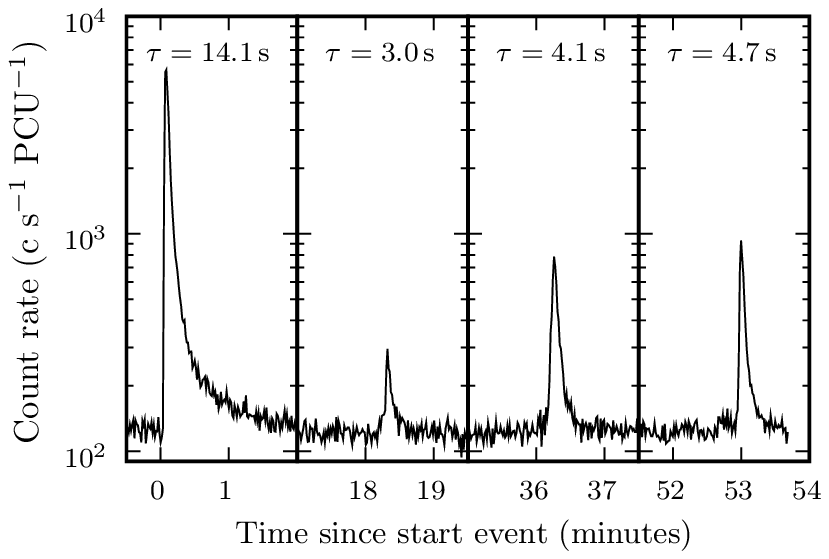}

\caption{\label{fig:quad}Quadruple burst from 4U~1636-53 as observed with
the \emph{RXTE} PCA on MJD 52286. Light curve at $2\,\mathrm{s}$
time resolution (\emph{top}) and zoomed in on each burst at $1\,\mathrm{s}$
time resolution (\emph{bottom}). For each burst we indicate the (longest)
exponential decay time $\tau$.}

\end{figure}
To illustrate the properties of LWT and SWT bursts, of which we have
shown the distributions using a large number of bursts from the catalog,
we consider the quadruple burst from 4U~1636-53 (Fig.~\ref{fig:quad}).
The four bursts occurred within $54\,\mathrm{minutes}$, and the time
between the burst onsets is $18.2\,\mathrm{min}$, $17.9\,\mathrm{min}$,
and $16.8\,\mathrm{min}$, respectively. As indicated in Fig.~\ref{fig:quad},
the first burst has by far the longest decay time $\tau$. The first
burst has a peak flux and fluence that is over seven times larger
than any of the three SWT bursts. The peak black-body temperature
($\mathrm{k}T$) of the four bursts is, respectively, $(1.67\pm0.02)\,\mathrm{keV}$,
$(1.06\pm0.06)\,\mathrm{keV}$, $(1.51\pm0.06)\,\mathrm{keV}$, and
$(1.50\pm0.04)\,\mathrm{keV}$: again the first burst has the highest
value. The combined net burst fluence of the quadruple event is $(2.93\pm0.05)\,10^{39}\mathrm{erg}$,
which is $35\%$ higher than the average fluence of single bursts
from this source. and $27\%$ higher than the fluence of the first
burst from the quadruplet. This means that at least $27\%$ of the
available fuel did not burn in the initial burst.

\section{Discussion}

We study the short recurrence time behavior in a large sample of bursts
from multiple sources. We use a preliminary version of the MINBAR
burst catalog, containing $3387$ Type I X-ray bursts from $65$ sources.
$15$ sources exhibit bursts with recurrence times less then one hour:
SWT bursts. The short recurrence times do not allow for the accretion
of the hydrogen and helium that is burned during the burst, which
means that it must have been accreted before the previous burst. For
example, the burst fluences from the quadruple event we observe from
4U~1636-53 indicate that at least $27\%$ of the accreted fuel did
not burn in the first burst. This is in contradiction with current
one-dimensional multi-zone models, which predict that during a flash
over $90\%$ of the available fuel is burned (e.g., \citealt{Woosley2004}).
The aim of this study is to provide a comprehensive observational
assessment of the SWT behavior.

\subsection{Temperature and energetics}

We perform time resolved spectroscopy of the bursts, and find that
SWT bursts are on average less bright and reach a lower black-body
temperature at the peak than the LWT bursts. The fluence of SWT bursts
is on average lower, but the combined bursts in a multiple-burst event
are as energetic as the most energetic single bursts. This is in agreement
with the \emph{XMM-Newton} observations of EXO~0748-676 analyzed
by \citet{Boirin2007}. In contrast to that investigation, however,
we find that the distributions of these quantities for single and
for first bursts of multiple-burst events are not compatible according
to Kolmogorov-Smirnov tests. It is possible that our improved statistics
allows us to see a disparity that previously went unnoticed.

\subsection{Multiple surface regions vs. multiple layers}

Several ideas have been put forward to explain SWT bursts in an attempt
to answer the two main questions: how to preserve fuel during a burst
for the next burst, and how to reignite this fuel on a time scale
of approximately ten minutes. Subsequent bursts with short recurrence
time may take place in different regions of the neutron star surface.
The accreted matter may be confined by a magnetic dipole field to
the poles of the neutron star, or perhaps the burning front of a burst
is stalled at the equator if the inflow of matter from the accretion
disk is particularly strong there. This would provide an explanation
for double bursts, but the triple and quadruple bursts that we observe
would require a more complicated configuration of the magnetic field.
Furthermore, \citet{Boirin2007} find for EXO~0748-676 that there
is no evidence for a difference in the X-ray emitting region during
the different bursts of multiple-burst events. Also, they find indications
that SWT bursts burn a fuel mixture with a lower hydrogen content,
which would not be the case if different regions of pristine accreted
material are burned. We, therefore, favor the scenario where SWT bursts
take place in different layers on top of each other (\citealt{1636:fujimoto87apj}).
For this to work, the thermonuclear burning during the flash must
be halted before the hydrogen and helium is depleted.

\subsection{No SWT bursts from UCXBs}

We observe no SWT bursts from any of the $16$ (candidate) ultra-compact
sources from which in total $229$ bursts have been observed with
the PCA and WFCs. For this reason we excluded these sources from our
analyses. For the hydrogen-accreting sources we found an fraction
of $3.1\%$ of all bursts are SWT bursts. If we assume the same SWT
fraction for the most frequently bursting confirmed UCXB, \object{4U~1820-303},
we expect $1.7$ SWT bursts out of the $54$ bursts observed by the
PCA and WFCs. The Poisson probability of detecting no SWT bursts when
expecting $1.7$, is $0.18$. Taking into account the bursts from
all confirmed and candidate UCXBs, we expect $7$ SWT bursts, and
the probability of a non-detection is less then $10^{-3}$.

Based on its bursting behavior, 4U~1728-34 is a suspected UCXB (\citealt{Galloway2008catalog}).
If we include the $543$ observed by the PCA and WFCs from this source,
the expected number of SWT bursts is $23$, and the probability of
detecting none is $10^{-10}$.

\subsection{Thermonuclear burning processes}

No SWT bursts are observed from UCXBs, but only from sources that
are thought to accrete hydrogen-rich matter. This suggests that the
nuclear burning processes involving hydrogen, i.e., the hot CNO cycle,
the $\alpha$\textsl{p}-process, and the \textsl{rp}-process, are
important for creating SWT bursts. The \textsl{rp}-process is a series
of proton captures and $\beta$-decays, that creates heavy isotopes
with mass numbers up to approximately $100$ (\citealt{Schatz2001}).
In this reaction chain there might be a nuclear waiting point with
the correct time scale to interrupt and reignite the thermonuclear
burning, for example the time scale for spontaneous $\beta$-decay
of an isotope. We find, however, a broad distribution of short recurrence
times $t_{\mathrm{recur}}$ (Fig.~\ref{fig:twait_raw}), which argues
against a single waiting point in the reaction chain (see also \citealt{Boirin2007}).
Note that we do not detect SWT bursts from all hydrogen-rich accretors,
the two frequent bursters KS~1731-26 and GS~1826-24 being the best
examples. This means that merely accreting hydrogen is not enough
to produce SWT bursts.

The decay profile of an X-ray burst is shaped by two processes. First
there is radiative cooling on a thermal time scale. For normal bursts,
which ignite at a typical column depth of $y\simeq10^{8}\,\mathrm{g\, cm^{2}}$,
this time scale is $\tau_{\mathrm{therm}}\simeq10\,\mathrm{s}$. A
second process, that slows down the decay, is prolonged thermonuclear
burning through the \textsl{rp}-process, which lasts up to approximately
$100\,\mathrm{s}$ (\citealt{Schatz2001}). For EXO~0748-676 \citet{Boirin2007}
found SWT bursts to lack the second slower decay component, while
the first bursts clearly exhibit a two-component exponential decay.
We confirm that this holds true for the other sources with SWT bursts
as well. This supports the conclusion by \citet{Boirin2007} that
follow-up bursts must occur in a layer with a significantly reduced
hydrogen content.

\citet{Zand2009} found observations where the burst decay can be
followed for several thousands of seconds. They explain the long tail
as due to the cooling of a deeper layer below the bursting layer,
that is heated by the burst. One of these long tails is detected for
the first burst in a triple event of EXO~0748-676. The tail continues
to decay uninterrupted while the second and third bursts occur. This
is consistent with the idea that SWT bursts occur in a layer above
the ignition depth where the first burst occurs. Taking into account
that SWT bursts are less energetic, the deeper layer would not be
heated substantially by the SWT bursts and continues to cool.

\subsection{Mass accretion rate dependence}

SWT bursts are observed over the entire range of mass accretion rates
$\dot{M}$ where LWT bursts are observed. For some individual sources,
however, the SWT bursts occur in a smaller $\dot{M}$ interval than
LWT bursts. This is supported by the position in the color-color diagram
where SWT bursts are observed (Sect.~\ref{sub:Accretion-rate-dependence};
\citealt{Galloway2008catalog}). Other frequent bursters exhibit no
SWT bursts at all, even though the ranges of $\dot{M}$ we observe
from sources with and without SWT bursts overlap. Because of the large
uncertainty in converting flux to accretion rate, the precise overlap
is uncertain. 

At low accretion rates SWT recurrence times $t_{\mathrm{recur}}$
are mostly restricted to $3.8\,\mathrm{min}\lesssim t_{\mathrm{recur}}\lesssim40\,\mathrm{min}.$
Above approximately $0.05\,\mathrm{\dot{M}_{\mathrm{Edd}}}$, where
the Eddington limited mass accretion rate $\mathrm{\dot{M}}_{\mathrm{Edd}}$
corresponds to a persistent luminosity of $\mathrm{L_{\mathrm{Edd}}}=2\cdot10^{38}\,\mathrm{erg\, s^{-1}}$
for hydrogen-accreting sources, recurrence times occur also in the
range $40\,\mathrm{min}\lesssim t_{\mathrm{recur}}\lesssim60\,\mathrm{min}.$
At $0.05\,\mathrm{\dot{M}_{\mathrm{Edd}}}$ a transition between two
bursts regimes is predicted by \citet{Fujimoto1981} (see also \citealt{Bildsten1998}).
For lower accretion rates all accreted hydrogen burns in a stable
manner, and the burst ignites in a hydrogen-poor layer. At higher
rates there is no time to burn all hydrogen, and the burst ignites
in a layer containing a substantial fraction of both hydrogen and
helium. It may be that the latter regime allows for short recurrence
times as long as an hour, while the former regime does not.

\subsection{Rotation and mixing}

The spin frequency $\nu_{\mathrm{spin}}$ is not known for most accreting
neutron stars, as it requires the observation of X-ray pulsations
or burst oscillations (e.g., \citealt{Galloway2008}). For five sources
with short recurrence times $\nu_{\mathrm{spin}}$ is known: all five
are fast spinning neutron stars with $\nu_{\mathrm{spin}}\gtrsim500\,\mathrm{Hz}$
(Table~\ref{tab:Overview-of-16}). The fast rotation could be required
for the occurrence of multiple-burst events. It induces rotational
instabilities, for example shear instabilities (e.g, \citealt{Fujimoto1988A&A}),
and instabilities due to a rotationally induced magnetic field (\citealt{Spruit2002}),
that mix the neutron star envelope on a time scale of approximately
ten minutes (\citealt{Piro2007,Keek2009}). If the thermonuclear burning
during a flash is halted before it reaches higher layers, the hydrogen
and helium in those layers will be mixed down. On a ten minute time
scale it reaches the depth where temperature and density are sufficiently
high to create the thermonuclear runaway for the next burst.

At accretion rates higher than $0.05\,\dot{M}_{\mathrm{Edd}}$, we
find bursts with short recurrence times as long as an hour. We mentioned
that this may be related to the transition to a different burst regime.
The time scale for rotational mixing depends strongly on the thermal
and compositional profile of the neutron star envelope (e.g., \citealt{Heger2000}
for the case of massive stars), which may vary for different burning
regimes or even different bursts. This could provide an explanation
for the spread in the observed recurrence times. Further theoretical
study is necessary to better understand this.

The occurrence of multiple-burst events in sources with high rotation
rates has consequences for the hypothesis that strong magnetic fields
at the neutron star surface contain accreted matter at the poles.
This would explain multiple bursts as caused by the burning of different
magnetically-confined patches at the poles. The presence of a strong
magnetic field, however, would allow for the transportation of angular
momentum away from the neutron star, causing it to spin slower. The
observations of short recurrence time bursts preferentially at high
$\nu_{\mathrm{spin}}$ seems in contradiction with this, which disfavors
the magnetic-confinement scenario.

KS~1731-260 spins at a high frequency of $524\,\mathrm{Hz}$, accreted
hydrogen-rich material, and exhibited many bursts ($369$ in MINBAR).
No short recurrence times were observed. Therefore, while a high rotation
rate may support the occurrence of SWT bursts, it is not possible
to discriminate between sources with and without SWT bursts based
on this property alone. Possibly a combination of fast rotation and
a mass accretion rate within a certain range (see previous section)
are required for short recurrence times.

\subsection{Frequency of SWT bursts}

We investigate the SWT fraction: the number of bursts that have a
short recurrence time with respect to the total number of observed
bursts. We determine this fraction, at the persistent luminosities
where SWT bursts are observed, for three frequent bursters with SWT
bursts: EXO~0748-676, 4U~1636-53, and 2S~1742-294. \emph{XMM-Newton}
and \emph{Chandra} observations of EXO~0748-676 find an SWT fraction
of $(30\pm5)\%$. Our burst sample is obtained from \emph{RXTE} PCA
and \emph{BeppoSAX} WFC observations, which contain data gaps due
to Earth occultations and due to the South-Atlantic Anomaly. Monte-Carlo
simulations show that the data gaps reduce an SWT fraction of $30\%$
to $20\%$. An additional problem is the fact that the WFCs are less
sensitive to fainter bursts than the PCA or the instruments on \emph{XMM-Newton}
and \emph{Chandra}. Especially for EXO~0748-676 we find a much lower
SWT fraction than from the \emph{XMM} and \emph{Chandra} observations.
Taking only the PCA bursts into account, we find SWT fractions of
$(11\pm6)\%$ for EXO~0748-676, $(13\pm4)\%$ for 4U~1636-53, and
$(23\pm6)\%$ for 2S~1742-294, all of which are within $1.5\sigma$
from the SWT fractions we obtain from Monte Carlo simulations, with
an initial SWT fraction of $30\%$. Therefore, in the range of mass
accretion rates where SWT bursts occur, approximately $30\%$ of the
bursts have a short recurrence time.

\section{Conclusions}

We studied thermonuclear bursts with short recurrence times (SWT)
using a large catalog of bursts from multiple sources observed with
the \emph{RXTE} PCA and the \emph{BeppoSAX} WFCs. The short recurrence
times are of insufficient duration to accrete the fuel that burns
in an SWT burst. Bursts are seen to occur in events of up to four
bursts: double, triple and quadruple bursts. We report the shortest
recurrence time ever found: $3.8$~minutes. We confirm the result
of \citet{Boirin2007}, that SWT bursts are on average less bright,
cooler, and less energetic than LWT bursts. The decay profiles of
SWT bursts lack the longer decay component from the \textsl{rp}-process,
suggesting that SWT bursts take place in a hydrogen-depleted layer.

Some sources exhibit short recurrence times at all values of the mass
accretion rate where normal bursts occur, while for others SWT bursts
are limited to a smaller range of accretion rate. In this range the
fraction of bursts with a short recurrence time is consistent with
$30\%$. Two frequent bursters that likely accrete hydrogen-rich matter
do not show any SWT bursts.

Only the hydrogen-accreting neutron stars in our catalog exhibit SWT
bursts. This suggests that the hydrogen-burning processes are responsible
for the incomplete burning of the available hydrogen and helium during
bursts. The mechanism for halting the burning is still unknown. It
will require further theoretical modeling of hydrogen-accreting neutron
star envelopes to resolve this issue.

As far as we know the spin of the sources with SWT bursts, they are
all fast rotators. This indicates that rotational mixing can be responsible
for ignition of follow-up bursts on a time scale of approximately
$10$~minutes (\citealt{Piro2007,Keek2009}). The number of SWT sources
with known spin is small. Measurements of the spin frequency for more
sources and further model studies of rotational mixing will better
constrain this reignition scenario.

\acknowledgements{LK is supported by the Joint Institute for Nuclear Astrophysics (JINA;
grant PHY02-16783), a National Science Foundation Physics Frontier
Center. AH acknowledges support from the DOE Program for Scientific
Discovery through Advanced Computing (SciDAC; DE-FC02-09ER41618) and
and by the US Department of Energy under grant DE-FG02-87ER40328.}

\bibliographystyle{apj}
\bibliography{apj-jour,keek0518}

\begin{thebibliography}{64}
\expandafter\ifx\csname natexlab\endcsname\relax\def\natexlab#1{#1}\fi

\bibitem[{{Anders} \& {Ebihara}(1982)}]{Anders1982}
{Anders}, E., \& {Ebihara}, M. 1982, \gca, 46, 2363

\bibitem[{{Aoki} {et~al.}(1992){Aoki}, {Dotani}, {Ebisawa}, {Itoh}, {Makino},
  {Nagase}, {Takeshima}, {Mihara}, \& {Kitamoto}}]{aoki92pasj}
{Aoki}, T., {et~al.} 1992, \pasj, 44, 641

\bibitem[{{Belian} {et~al.}(1976){Belian}, {Conner}, \& {Evans}}]{1976Belian}
{Belian}, R.~D., {Conner}, J.~P., \& {Evans}, W.~D. 1976, \apjl, 206, L135

\bibitem[{{Bildsten}(1998)}]{Bildsten1998}
{Bildsten}, L. 1998, in NATO ASIC Proc. 515: The Many Faces of Neutron Stars.,
  ed. R.~{Buccheri}, J.~{van Paradijs}, \& A.~{Alpar}, 419

\bibitem[{{Boella} {et~al.}(1997){Boella}, {Butler}, {Perola}, {Piro},
  {Scarsi}, \& {Bleeker}}]{1997Boella}
{Boella}, G., {Butler}, R.~C., {Perola}, G.~C., {Piro}, L., {Scarsi}, L., \&
  {Bleeker}, J.~A.~M. 1997, \aaps, 122, 299

\bibitem[{{Boirin} {et~al.}(2007){Boirin}, {Keek}, {M{\'e}ndez}, {Cumming},
  {in't Zand}, {Cottam}, {Paerels}, \& {Lewin}}]{Boirin2007}
{Boirin}, L., {Keek}, L., {M{\'e}ndez}, M., {Cumming}, A., {in't Zand},
  J.~J.~M., {Cottam}, J., {Paerels}, F., \& {Lewin}, W.~H.~G. 2007, \aap, 465,
  559

\bibitem[{{Cornelisse} {et~al.}(2003){Cornelisse}, {in~'t~Zand}, {Verbunt},
  {Kuulkers}, {Heise}, {den Hartog}, {Cocchi}, {Natalucci}, {Bazzano}, \&
  {Ubertini}}]{Cornelisse2003}
{Cornelisse}, R., {et~al.} 2003, \aap, 405, 1033

\bibitem[{{Fender} {et~al.}(2005){Fender}, {Belloni}, \& {Gallo}}]{Fender2005}
{Fender}, R., {Belloni}, T., \& {Gallo}, E. 2005, \apss, 300, 1

\bibitem[{{Fujimoto}(1988)}]{Fujimoto1988A&A}
{Fujimoto}, M.~Y. 1988, \aap, 198, 163

\bibitem[{{Fujimoto} {et~al.}(1981){Fujimoto}, {Hanawa}, \&
  {Miyaji}}]{Fujimoto1981}
{Fujimoto}, M.~Y., {Hanawa}, T., \& {Miyaji}, S. 1981, \apj, 247, 267

\bibitem[{{Fujimoto} {et~al.}(1987){Fujimoto}, {Sztajno}, {Lewin}, \& {van
  Paradijs}}]{1636:fujimoto87apj}
{Fujimoto}, M.~Y., {Sztajno}, M., {Lewin}, W.~H.~G., \& {van Paradijs}, J.
  1987, \apj, 319, 902

\bibitem[{{Fushiki} \& {Lamb}(1987)}]{Fushiki1987ApJ}
{Fushiki}, I., \& {Lamb}, D.~Q. 1987, \apjl, 323, L55

\bibitem[{{Galloway}(2008)}]{Galloway2008}
{Galloway}, D. 2008, in American Institute of Physics Conference Series, Vol.
  983, 40 Years of Pulsars: Millisecond Pulsars, Magnetars and More, 510--518

\bibitem[{{Galloway} {et~al.}(2004{\natexlab{a}}){Galloway}, {Chakrabarty},
  {Cumming}, {Kuulkers}, {Bildsten}, \& {Rothschild}}]{Galloway2004AIPC}
{Galloway}, D.~K., {Chakrabarty}, D., {Cumming}, A., {Kuulkers}, E.,
  {Bildsten}, L., \& {Rothschild}, R. 2004{\natexlab{a}}, in American Institute
  of Physics Conference Series, Vol. 714, X-ray Timing 2003: Rossi and Beyond,
  ed. {P.~Kaaret, F.~K.~Lamb, \& J.~H.~Swank}, 266--272

\bibitem[{{Galloway} {et~al.}(2004{\natexlab{b}}){Galloway}, {Cumming},
  {Kuulkers}, {Bildsten}, {Chakrabarty}, \& {Rothschild}}]{1826:galloway04apj}
{Galloway}, D.~K., {Cumming}, A., {Kuulkers}, E., {Bildsten}, L.,
  {Chakrabarty}, D., \& {Rothschild}, R.~E. 2004{\natexlab{b}}, \apj, 601, 466

\bibitem[{{Galloway} {et~al.}(2009){Galloway}, {Lin}, {Chakrabarty}, \&
  {Hartman}}]{Galloway2009}
{Galloway}, D.~K., {Lin}, J., {Chakrabarty}, D., \& {Hartman}, J.~M. 2009,
  ArXiv e-prints

\bibitem[{{Galloway} {et~al.}(2008){Galloway}, {Muno}, {Hartman}, {Psaltis}, \&
  {Chakrabarty}}]{Galloway2008catalog}
{Galloway}, D.~K., {Muno}, M.~P., {Hartman}, J.~M., {Psaltis}, D., \&
  {Chakrabarty}, D. 2008, \apjs, 179, 360

\bibitem[{{Gold}(1968)}]{Gold1968}
{Gold}, T. 1968, \nat, 218, 731

\bibitem[{{Gottwald} {et~al.}(1986){Gottwald}, {Haberl}, {Parmar}, \&
  {White}}]{0748:gottwald86apj}
{Gottwald}, M., {Haberl}, F., {Parmar}, A.~N., \& {White}, N.~E. 1986, \apj,
  308, 213

\bibitem[{{Gottwald} {et~al.}(1987{\natexlab{a}}){Gottwald}, {Haberl},
  {Parmar}, \& {White}}]{0748:gottwald87apj}
---. 1987{\natexlab{a}}, \apj, 323, 575

\bibitem[{{Gottwald} {et~al.}(1987{\natexlab{b}}){Gottwald}, {Stella}, {White},
  \& {Barr}}]{Gottwald1987}
{Gottwald}, M., {Stella}, L., {White}, N.~E., \& {Barr}, P. 1987{\natexlab{b}},
  \mnras, 229, 395

\bibitem[{{Grindlay} {et~al.}(1976){Grindlay}, {Gursky}, {Schnopper},
  {Parsignault}, {Heise}, {Brinkman}, \& {Schrijver}}]{Grindlay1976}
{Grindlay}, J., {Gursky}, H., {Schnopper}, H., {Parsignault}, D.~R., {Heise},
  J., {Brinkman}, A.~C., \& {Schrijver}, J. 1976, \apjl, 205, L127

\bibitem[{{Gruber} {et~al.}(1996){Gruber}, {Blanco}, {Heindl}, {Pelling},
  {Rothschild}, \& {Hink}}]{Gruber1996}
{Gruber}, D.~E., {Blanco}, P.~R., {Heindl}, W.~A., {Pelling}, M.~R.,
  {Rothschild}, R.~E., \& {Hink}, P.~L. 1996, \aaps, 120, C641+

\bibitem[{{Hasinger} \& {van der Klis}(1989)}]{1989Hasinger}
{Hasinger}, G., \& {van der Klis}, M. 1989, \aap, 225, 79

\bibitem[{{Heger} {et~al.}(2007){Heger}, {Cumming}, {Galloway}, \&
  {Woosley}}]{Heger2007}
{Heger}, A., {Cumming}, A., {Galloway}, D.~K., \& {Woosley}, S.~E. 2007, \apjl,
  671, L141

\bibitem[{{Heger} {et~al.}(2000){Heger}, {Langer}, \& {Woosley}}]{Heger2000}
{Heger}, A., {Langer}, N., \& {Woosley}, S.~E. 2000, \apj, 528, 368

\bibitem[{{Homan} {et~al.}(2003){Homan}, {Wijnands}, \& {van den
  Berg}}]{Homan2003}
{Homan}, J., {Wijnands}, R., \& {van den Berg}, M. 2003, \aap, 412, 799

\bibitem[{{in~'t~Zand} {et~al.}(2005){in~'t~Zand}, {Cumming}, {van der Sluys},
  {Verbunt}, \& {Pols}}]{Zand2005}
{in~'t~Zand}, J.~J.~M., {Cumming}, A., {van der Sluys}, M.~V., {Verbunt}, F.,
  \& {Pols}, O.~R. 2005, \aap, 441, 675

\bibitem[{{in~'t~Zand} {et~al.}(2007){in~'t~Zand}, {Jonker}, \&
  {Markwardt}}]{intZand2007}
{in~'t~Zand}, J.~J.~M., {Jonker}, P.~G., \& {Markwardt}, C.~B. 2007, \aap, 465,
  953

\bibitem[{{in~'t~Zand} {et~al.}(2009){in~'t~Zand}, {Keek}, {Cumming}, {Heger},
  {Homan}, \& {M{\'e}ndez}}]{Zand2009}
{in~'t~Zand}, J.~J.~M., {Keek}, L., {Cumming}, A., {Heger}, A., {Homan}, J., \&
  {M{\'e}ndez}, M. 2009, \aap, 497, 469

\bibitem[{{in~'t~Zand} {et~al.}(2003){in~'t~Zand}, {Kuulkers}, {Verbunt},
  {Heise}, \& {Cornelisse}}]{Zand2003}
{in~'t~Zand}, J.~J.~M., {Kuulkers}, E., {Verbunt}, F., {Heise}, J., \&
  {Cornelisse}, R. 2003, \aap, 411, L487

\bibitem[{{in~'t~Zand} {et~al.}(2004){in~'t~Zand}, {Verbunt}, {Heise},
  {Bazzano}, {Cocchi}, {Cornelisse}, {Kuulkers}, {Natalucci}, \&
  {Ubertini}}]{2004ZandWFC}
{in~'t~Zand}, J.~J.~M., {et~al.} 2004, Nucl. Phys. Proc. Suppl., 132, 486

\bibitem[{{Jager} {et~al.}(1997){Jager}, {Mels}, {Brinkman}, {Galama},
  {Goulooze}, {Heise}, {Lowes}, {Muller}, {Naber}, {Rook}, {Schuurhof},
  {Schuurmans}, \& {Wiersma}}]{1997Jager}
{Jager}, R., {et~al.} 1997, \aaps, 125, 557

\bibitem[{{Jahoda} {et~al.}(2006){Jahoda}, {Markwardt}, {Radeva}, {Rots},
  {Stark}, {Swank}, {Strohmayer}, \& {Zhang}}]{Jahoda2006}
{Jahoda}, K., {Markwardt}, C.~B., {Radeva}, Y., {Rots}, A.~H., {Stark}, M.~J.,
  {Swank}, J.~H., {Strohmayer}, T.~E., \& {Zhang}, W. 2006, \apjs, 163, 401

\bibitem[{{Keek} {et~al.}(2006){Keek}, {in~'t~Zand}, \& {Cumming}}]{kee06}
{Keek}, L., {in~'t~Zand}, J.~J.~M., \& {Cumming}, A. 2006, \aap, 455, 1031

\bibitem[{{Keek} {et~al.}(2009){Keek}, {Langer}, \& {in~'t~Zand}}]{Keek2009}
{Keek}, L., {Langer}, N., \& {in~'t~Zand}, J.~J.~M. 2009, \aap, 502, 871

\bibitem[{{Kuulkers} {et~al.}(2003){Kuulkers}, {den Hartog}, {in~'t~Zand},
  {Verbunt}, {Harris}, \& {Cocchi}}]{Kuulkers2003}
{Kuulkers}, E., {den Hartog}, P.~R., {in~'t~Zand}, J.~J.~M., {Verbunt},
  F.~W.~M., {Harris}, W.~E., \& {Cocchi}, M. 2003, \aap, 399, 663

\bibitem[{{Kuulkers} {et~al.}(2002){Kuulkers}, {Homan}, {van der Klis},
  {Lewin}, \& {M{\'e}ndez}}]{2002Kuulkers}
{Kuulkers}, E., {Homan}, J., {van der Klis}, M., {Lewin}, W.~H.~G., \&
  {M{\'e}ndez}, M. 2002, \aap, 382, 947

\bibitem[{{Lamb} \& {Lamb}(1978)}]{Lamb1978}
{Lamb}, D.~Q., \& {Lamb}, F.~K. 1978, \apj, 220, 291

\bibitem[{{Lamb} {et~al.}(2009){Lamb}, {Boutloukos}, {Van Wassenhove},
  {Chamberlain}, {Lo}, {Clare}, {Yu}, \& {Miller}}]{Lamb2009}
{Lamb}, F.~K., {Boutloukos}, S., {Van Wassenhove}, S., {Chamberlain}, R.~T.,
  {Lo}, K.~H., {Clare}, A., {Yu}, W., \& {Miller}, M.~C. 2009, \apj, 706, 417

\bibitem[{{Lattimer} \& {Prakash}(2007)}]{Lattimer2007}
{Lattimer}, J.~M., \& {Prakash}, M. 2007, \physrep, 442, 109

\bibitem[{{Lewin} {et~al.}(1976){Lewin}, {Hoffman}, {Doty}, {Hearn}, {Clark},
  {Jernigan}, {Li}, {McClintock}, \& {Richardson}}]{lewin76mnras}
{Lewin}, W.~H.~G., {et~al.} 1976, \mnras, 177, 83P

\bibitem[{{Lewin} {et~al.}(1993){Lewin}, {van Paradijs}, \& {Taam}}]{Lewin1993}
{Lewin}, W.~H.~G., {van Paradijs}, J., \& {Taam}, R.~E. 1993, Space Science
  Reviews, 62, 223

\bibitem[{{Linares} {et~al.}(2009){Linares}, {Watts}, {Altamirano}, {Patruno},
  {Casella}, {Rea}, {Soleri}, {van der Klis}, {Wijnands}, {Belloni}, {Homan},
  \& {Mendez}}]{Linares2009ATel}
{Linares}, M., {et~al.} 2009, The Astronomer's Telegram, 1979, 1

\bibitem[{{Liu} {et~al.}(2007){Liu}, {van Paradijs}, \& {van den
  Heuvel}}]{Liu2007}
{Liu}, Q.~Z., {van Paradijs}, J., \& {van den Heuvel}, E.~P.~J. 2007, \aap,
  469, 807

\bibitem[{{Maraschi} \& {Cavaliere}(1977)}]{Maraschi1977}
{Maraschi}, L., \& {Cavaliere}, A. 1977, in Highlights in Astronomy, ed. E.~A.
  {M\"uller}, Vol.~4 (Reidel, Dordrecht), 127

\bibitem[{{Melatos} \& {Payne}(2005)}]{Melatos2005}
{Melatos}, A., \& {Payne}, D.~J.~B. 2005, \apj, 623, 1044

\bibitem[{{Morrison} \& {McCammon}(1983)}]{1983Morrison}
{Morrison}, R., \& {McCammon}, D. 1983, \apj, 270, 119

\bibitem[{{Muno} {et~al.}(2002){Muno}, {Chakrabarty}, {Galloway}, \&
  {Psaltis}}]{Muno2002}
{Muno}, M.~P., {Chakrabarty}, D., {Galloway}, D.~K., \& {Psaltis}, D. 2002,
  \apj, 580, 1048

\bibitem[{{Murakami} {et~al.}(1980){Murakami}, {Inoue}, {Koyama}, {Makishima},
  {Matsuoka}, {Oda}, {Ogawara}, {Ohashi}, {Shibazaki}, {Tanaka}, {Hayakawa},
  {Kunieda}, {Makino}, {Masai}, {Nagase}, {Tawara}, {Miyamoto}, {Tsunemi},
  {Yamashita}, \& {Kondo}}]{1608:murakami80pasj}
{Murakami}, T., {et~al.} 1980, \pasj, 32, 543

\bibitem[{{Piro} \& {Bildsten}(2007)}]{Piro2007}
{Piro}, A.~L., \& {Bildsten}, L. 2007, \apj, 663, 1252

\bibitem[{{Schatz} {et~al.}(2001){Schatz}, {Aprahamian}, {Barnard}, {Bildsten},
  {Cumming}, {Ouellette}, {Rauscher}, {Thielemann}, \& {Wiescher}}]{Schatz2001}
{Schatz}, H., {et~al.} 2001, Physical Review Letters, 86, 3471

\bibitem[{{Smith} {et~al.}(1997){Smith}, {Morgan}, \& {Bradt}}]{Smith1997}
{Smith}, D.~A., {Morgan}, E.~H., \& {Bradt}, H. 1997, \apjl, 479, L137+

\bibitem[{{Spruit}(2002)}]{Spruit2002}
{Spruit}, H.~C. 2002, \aap, 381, 923

\bibitem[{{Strohmayer} \& {Bildsten}(2006)}]{Strohmayer2006}
{Strohmayer}, T., \& {Bildsten}, L. 2006, {New views of thermonuclear bursts}
  (Compact stellar X-ray sources), 113--156

\bibitem[{{Strohmayer} {et~al.}(1998){Strohmayer}, {Zhang}, {Swank}, {White},
  \& {Lapidus}}]{Strohmayer1998}
{Strohmayer}, T.~E., {Zhang}, W., {Swank}, J.~H., {White}, N.~E., \& {Lapidus},
  I. 1998, \apjl, 498, L135+

\bibitem[{{Taam} {et~al.}(1993){Taam}, {Woosley}, {Weaver}, \&
  {Lamb}}]{Taam1993}
{Taam}, R.~E., {Woosley}, S.~E., {Weaver}, T.~A., \& {Lamb}, D.~Q. 1993, \apj,
  413, 324

\bibitem[{{van Paradijs} {et~al.}(1988){van Paradijs}, {Penninx}, \&
  {Lewin}}]{Paradijs1988}
{van Paradijs}, J., {Penninx}, W., \& {Lewin}, W.~H.~G. 1988, \mnras, 233, 437

\bibitem[{{Wallace} \& {Woosley}(1981)}]{Wallace1981}
{Wallace}, R.~K., \& {Woosley}, S.~E. 1981, \apjs, 45, 389

\bibitem[{{Wijnands} {et~al.}(2002){Wijnands}, {Muno}, {Miller}, {Franco},
  {Strohmayer}, {Galloway}, \& {Chakrabarty}}]{1658:wijnands02apj}
{Wijnands}, R., {Muno}, M.~P., {Miller}, J.~M., {Franco}, L.~M., {Strohmayer},
  T., {Galloway}, D., \& {Chakrabarty}, D. 2002, \apj, 566, 1060

\bibitem[{{Wijnands} {et~al.}(2001){Wijnands}, {Strohmayer}, \&
  {Franco}}]{Wijnands2001}
{Wijnands}, R., {Strohmayer}, T., \& {Franco}, L.~M. 2001, \apjl, 549, L71

\bibitem[{{Woosley} {et~al.}(2004){Woosley}, {Heger}, {Cumming}, {Hoffman},
  {Pruet}, {Rauscher}, {Fisker}, {Schatz}, {Brown}, \&
  {Wiescher}}]{Woosley2004}
{Woosley}, S.~E., {et~al.} 2004, \apjs, 151, 75

\bibitem[{{Woosley} \& {Taam}(1976)}]{Woosley1976}
{Woosley}, S.~E., \& {Taam}, R.~E. 1976, \nat, 263, 101

\bibitem[{{Zhang} {et~al.}(1998){Zhang}, {Jahoda}, {Kelley}, {Strohmayer},
  {Swank}, \& {Zhang}}]{Zhang1998}
{Zhang}, W., {Jahoda}, K., {Kelley}, R.~L., {Strohmayer}, T.~E., {Swank},
  J.~H., \& {Zhang}, S.~N. 1998, \apjl, 495, L9+

\end{thebibliography}

\appendix{}

\section{Chandra observations of EXO~0748-676\label{sec:Chandra-observations-of}}

\begin{figure*}
\includegraphics[clip]{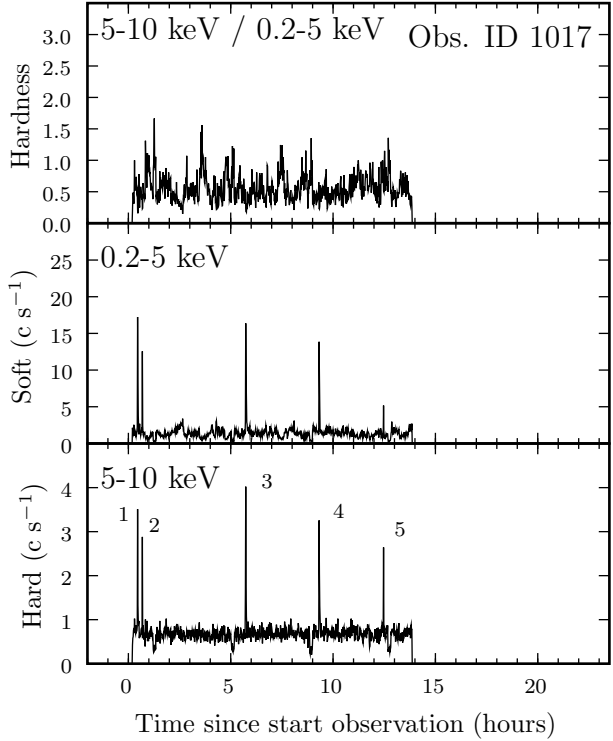}

\caption{\label{fig:Chandra-light-curves}\emph{Chandra} light curves of EXO
0748-676 as observed in 2001 and 2003. We show the hard and soft energy
bands as well as the hardness ratio. The observation id is indicated.
In the hard band we number the bursts. Figure continues on next page.}

\end{figure*}

\emph{Chandra} ACIS-S observations with High-Energy Transmission Grating
(HETG) in 2001 and 2003 of EXO~0748-676. We use the level 2 event
files prepared by the Chandra Data Archive and select events from
a circle centered on the source and two bands overlapping the signal
from the gratings. Using the CIAO software package (version 4.1),
we extract light curves in the $0.2-5\,\mathrm{keV}$ and $5-10\,\mathrm{keV}$
energy bands at $60\,\mathrm{s}$ time resolution, and calculate the
hardness ratio by taking the ratio of the count rates in the hard
and soft bands (Fig.~\ref{fig:Chandra-light-curves}). The light
curves are similar to the \emph{XMM-Newton} observations performed
in 2001 and 2003. Eclipses are present at the binary period of $3.8\,\mathrm{h}$.
Especially in the data from 2003, dipping is present in the soft band.
We, therefore, use the hard band to locate bursts. We find a total
of $41$ bursts; $11$ have a short recurrence time. The burst events
are divided into $20$ singles, $9$ doubles, and $1$ triple. The
SWT fraction is $(27\pm9)\%$. Note that burst $22$ of observation
4573 appears anomalously long. This is caused by a raised detector
background level during a few hundred seconds after the burst, combined
with the bin size of $60\,\mathrm{s}$ we used for the figure.

\begin{figure*}
\includegraphics[clip]{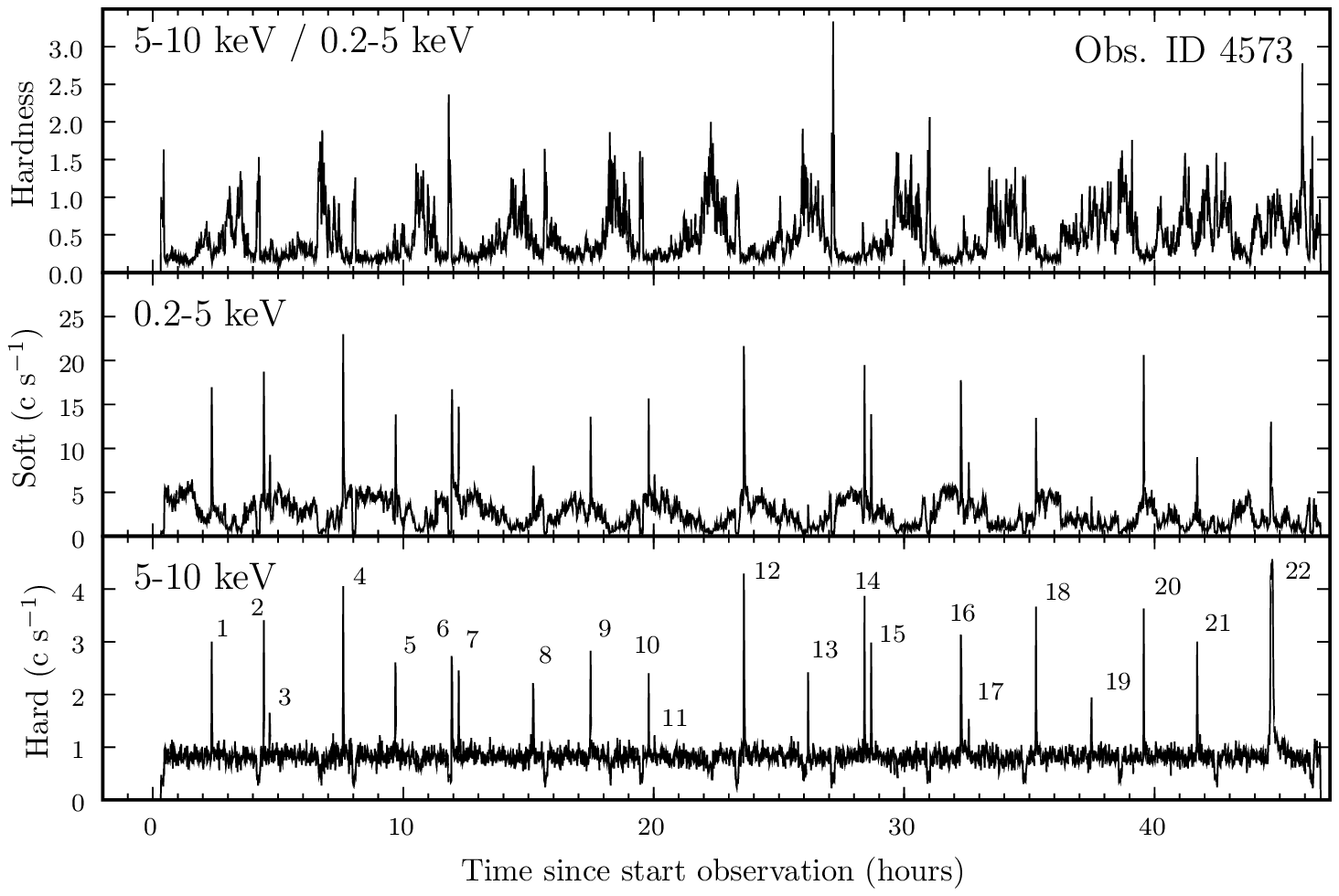}

\includegraphics[clip]{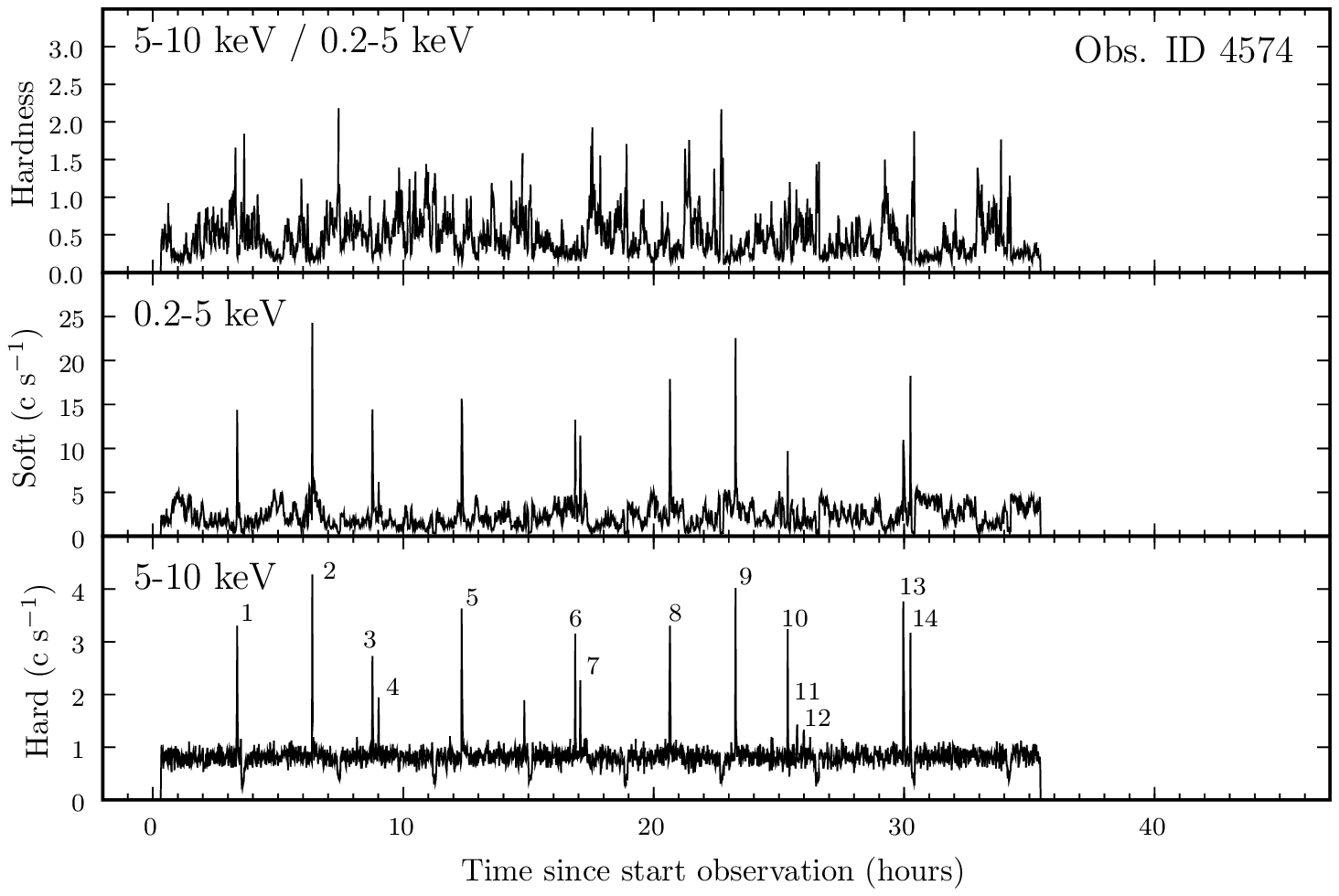}

\textsc{Fig.~\ref{fig:Chandra-light-curves} continued.}
\end{figure*}

\end{document}